\documentclass[numberedappendix, appendixfloats]{emulateapj}
\usepackage{natbib}
\usepackage{color}
\usepackage{comment}

\shorttitle{Photoevaporating Protoplanetary Disks}

\shortauthors{Nakatani {\em et al.}}

\newcommand{\HIImath}{\text{H{\cal II}}}
\newcommand{\HII}{\ion{H}{2}}

\newcommand{\OI}{\ion{O}{1}}

\newcommand{\e}[1]{\times 10^{#1}}

\newcommand{\fref}[1]{FIG.~\ref{#1}}
\newcommand{\tref}[1]{TABLE~\ref{#1}}
\newcommand{\eqnref}[1]{Eq.~(\ref{#1})} 
\newcommand{\secref}[1]{Section \ref{#1}}
\newcommand{\appref}[1]{Appendix \ref{#1}}

\newcommand{\braket}[1]{\left( {#1} \right)}

\newcommand{\cm}[1]{\,{\rm cm^{#1}}}
\newcommand{\AU}{\,{\rm au}}

\newcommand{\megayr}{\,{\rm Myr}}
\newcommand{\kms}{{\rm \,km\,s^{-1}}}
\newcommand{\eV}{{\rm \,eV}}

\newcommand{\keV}{\,{\rm keV}}
\newcommand{\myr}{\, M_{\odot}\, {\rm yr}^{-1}}
\newcommand{\unit}[2]{\,{\rm #1}^{#2}}
\newcommand{\Kelvin}{{\rm \, K}}

\newcommand{\abn}[1]{y_{\text{\rm #1}}}		
\newcommand{\col}[1]{N_{\text{\rm #1}}}		
\newcommand{\metal}{Z}					
\newcommand{\nh}{n_{\text{\rm H}}}			
\newcommand{\nspe}[1]{n_{\text{\rm #1}}}		
\newcommand{\smetal}{Z_\odot}			
\newcommand{\taux}{\tau_{\text{X}}}			
\newcommand{\Phih}{\Phi_{\text{H}}}			
\newcommand{\primaryioni}{\xi_\text{H, prim}}	
\newcommand{\secondioni}{ \xi_\text{H, sec}}	
\newcommand{\primary}[1]{\xi_\text{#1, prim}}	
\newcommand{\second}[1]{ \xi_\text{#1, sec}}	
\newcommand{\dgratio}{\mathscr{DG}}		
\newcommand{\Emax}{E_\text{max}}			
\newcommand{\Emin}{E_\text{min}}			

\newcommand{\crsc}[1]{\sigma_{\text{#1}}}	
\newcommand{\mdotph}{\dot{M}_\text{ph}}	
\newcommand{\vp}{v_\text{p}}				

\usepackage{mhchem, bm}
\usepackage{mathrsfs}	

\begin{document}

\title{Radiation hydrodynamics simulations of photoevaporation
of protoplanetary disks II: Metallicity dependence of UV and X-ray photoevaporation
}

\author{Riouhei~Nakatani\altaffilmark{1},
Takashi~Hosokawa\altaffilmark{2},
Naoki~Yoshida\altaffilmark{1,3,4},
Hideko~Nomura\altaffilmark{5},
and
Rolf Kuiper\altaffilmark{6}
}

\altaffiltext{1}{Department of Physics, School of Science, The University of Tokyo, 7-3-1 Hongo, Bunkyo, Tokyo 113-0033, Japan}

\altaffiltext{2}{Department of Physics, Kyoto University, Sakyo-ku, Kyoto, 606-8502, Japan}

\altaffiltext{3}{Kavli Institute for the Physics and Mathematics of the Universe (WPI),
       UT Institute for Advanced Study, The University of Tokyo, Kashiwa, Chiba 277-8583, Japan}

\altaffiltext{4}{Research Center for the Early Universe (RESCEU), 
School of Science, The University of Tokyo, 7-3-1 Hongo, Bunkyo, Tokyo 113-0033, Japan}
\altaffiltext{5}{Department of Earth and Planetary Sciences, Tokyo Institute of Technology, 2-12-1 Ookayama, Meguro, Tokyo, 152-8551, Japan}

\altaffiltext{6}{Institute of Astronomy and Astrophysics, University of T\"ubingen, Auf der Morgenstelle 10, D-72076 T\"ubingen, Germany}

\email{r.nakatani@utap.phys.s.u-tokyo.ac.jp}

\begin{abstract} 

We perform a suite of radiation hydrodynamics simulations 
of photoevaporating disks with varying 
the metallicity in a wide range of $10^{-3} ~\smetal \leq \metal \leq 10^{0.5} ~\smetal $.
We follow the disk evolution for over $\sim 5000$ years 
by solving hydrodynamics, radiative transfer, and non-equilibrium chemistry.
Our chemistry model is updated from the first paper of this series
by adding X-ray ionization and heating.
We study the metallicity dependence of the disk photoevaporation rate
and examine the importance of X-ray radiation.
In the fiducial case with solar metallicity,
including the X-ray effects does not significantly increase 
the photoevaporation rate when compared to the case with ultra-violet (UV) radiation only.
At sub-solar metallicities in the range of $\metal \gtrsim 10^{-1.5} \,\smetal $,
the photoevaporation rate increases as metallicity 
decreases owing to the reduced opacity of the disk medium.
The result is consistent with the observational trend that disk
lifetimes are shorter in low metallicity environments.
Contrastingly, the photoevaporation rate decreases at
even lower metallicities of $\metal \lesssim 10^{-1.5} \,\smetal $,
because dust-gas collisional cooling remains efficient
compared to far UV photoelectric heating
whose efficiency depends on metallicity.
The net cooling in the interior of the disk suppresses the photoevaporation. 
However, adding X-ray radiation significantly increases the photoevaporation rate,
especially at $\metal \sim 10^{-2}\, \smetal$.
Although the X-ray radiation itself does not drive strong photoevaporative flows,
X-rays penetrate deep into the neutral region in the disk,
increase the ionization degree there, and
reduce positive charges of grains.
Consequently, the effect of photoelectric heating by far UV radiation
is strengthened by the X-rays and enhances the disk photoevaporation.
\end{abstract}

\keywords{
protoplanetary disks -- stars: formation -- infrared: planetary systems 
-- stars: pre-main-sequence -- ultraviolet: stars
          }

\section{INTRODUCTION}
\label{introduction}

Infrared observations of star-forming regions show that the fraction
of infrared excess, a signature of the existence of protoplanetary disks (PPDs),
decreases with the age of the system, suggesting a finite lifetime
of $3-6\megayr$ for the PPDs in solar metallicity environments
\citep{2001_Haisch,2007_Hernandez,2007_Meyer,2009_Mamajek,2010_Fedele,2014_Ribas}.
There are a few physical mechanisms proposed theoretically,
which predict the disk dispersal time of a few million years,
but the exact processes that drive destruction or evaporation of a PPD are poorly known.
Interestingly, recent observations suggest that PPDs in sub-solar metallicity environments
	may have significantly shorter lifetimes of
	$\lesssim 1 \megayr$ \citep{2009_Yasui,2010_Yasui,2016_Yasui_I,2016_Yasui_II}.

	Photoevaporation, a physical process with which
	outflows are excited
	by irradiation from the central star,
	is proposed as a promising
	disk dispersal mechanism 
	\citep{1994_Hollenbach,2001_Clarke,2004_Alexander,
	2004_Font,2009_Ercolano,2009_GortiHollenbach,2010_Owen}.
	Far-ultraviolet (FUV; $6\eV \lesssim h \nu \leq 13.6\eV$),
	extreme-ultraviolet (EUV; $13.6\eV \leq h\nu \lesssim 0.1\keV$),
	and X-rays ($0.1\keV \leq h\nu  \leq 10\keV$)
	can cause photoevaporation through different physical processes.
	FUV radiation heats the disk gas by 
	photoelectric heating and/or photo-pumping of \ce{H2} \citep{2017_Wang},
	whereas photoionization heating by
        EUV and X-ray radiation can also drive evaporative flows from a PPD.
        
        EUV radiation is mainly absorbed by hydrogen 
        atoms in the disk gas,
        and 
        the absorption cross section per hydrogen atom, 
        of the order of $\sim 10^{-17} \cm{2}$, 
        is much larger than those for FUV and X-ray radiation
        that are absorbed by dust grains and hydrogen/helium/heavy elements, respectively.
        Therefore, FUV and X-ray radiation penetrate the deep interior of a disk
        with column densities of $\col{H} \sim 10^{21} \cm{-2}$, whereas
        EUV radiation effectively heats lower density regions close to the disk surfaces 
	with column densities of $\col{H} \sim 10^{19} - 10^{20} \cm{-2} $.
        
        Analytic models and numerical simulations suggest that EUV-driven photoevaporative flows
        originating from low-density regions
        (disk surfaces) yield a mass-loss rate of $\sim 10^{-10} - 10^{-9} \myr$
	\citep{1994_Hollenbach,2013_Tanaka}. It is smaller by a factor of ten to
        even hundreds than FUV- and X-ray-driven photoevaporation rates
	\citep[][hereafter Paper I]
	{2009_GortiHollenbach,2009_Ercolano,2010_Owen,2018_Nakatani}.	 
	Recent studies on photoevaporation of a PPD by FUV and X-ray radiation
        show rather diverse results, and the relative importance of FUV and X-rays is under debate 
	\citep{2009_GortiHollenbach, 2009_Ercolano, 2010_Owen, 2012_Owen, 2015_Gorti}.
	Interestingly, 
	\cite{2010_ErcolanoClarke} 
	show that X-ray photoevaporation is more efficient
	with lower metallicities
	owing to the reduced opacity effect.
        Unfortunately, these previous studies adopt
	different stellar models, disk models, and even different sets of chemistry,
        and thus one cannot compare the results directly.
        Furthermore, simplified assumptions are often made
	such as hydrostatic disk structure and/or radiative equilibrium,
        which degrades the reality of the calculations when considering
        the actual, dynamic evolution of a PPD.
	In order to examine the effect of FUV and X-ray radiation on PPD photoevaporation,
	it is necessary to perform hydrodynamics simulations 
        with all of the above physical processes included self-consistently.

	Two recent studies,
	\cite{2017_Wang} and our Paper I,
	use hydrodynamics simulations with radiative transfer and non-equilibrium chemistry 
	to follow the disk photoevaporation around a solar-type star.
	Both studies conclude that FUV photons effectively drive photoevaporation,
	although there are a few differences regarding the most effective heating process.
	In Paper I, 
	we investigate metallicity dependence of UV photoevaporation rates.
	We conclude that the FUV-driven photoevaporation rate 
	increases with decreasing metallicity 
	for $10^{-0.5} ~\smetal \lesssim \metal \lesssim 10~\smetal$.
	We also find that photoelectric heating due to FUV becomes inefficient 
	as metallicity decreases, 
	compared with dust-gas collisional cooling.
	This reduced FUV heating lowers the temperatures of the neutral region.
	For $\metal \lesssim 10^{-1.5} \, \smetal$,
	the neutral region temperatures are too low for photoevaporative flows to be excited.
	Only EUV-driven photoevaporative flows contribute to the mass loss in this case,
	and thus the photoevaporation rates
	are smaller by about an order of magnitude
	than those with $\metal \gtrsim 10^{-1}\, \smetal$.
	It is worth mentioning that the simulations of \cite{2017_Wang}, which incorporate
        X-ray heating, show that the X-ray radiation itself is not the primary
	cause of photoevaporation.
        
        In the present study,
	we perform a suite of simulations of photoevaporating protoplanetary disks 
	with various metallicities $10^{-3}\, \smetal \leq \metal \leq 10^{0.5}\, \smetal$.
	Our simulations incorporate X-ray heating and ionization
        coupled with our chemistry model of Paper I.
	We examine how X-ray radiation affects the disk photoevaporation rate,
	and determine the relative importance of FUV and X-ray in the process of photoevaporation.
	Also, we investigate
	the metallicity dependence of photoevaporation rates 
	due to both UV and X-ray.

	The paper is organized as follows.
	In \secref{sec:method}, 
	we present the methods of our simulations.
	In \secref{sec:result},
	we discuss the simulation results.
	A final discussion and a summary are given 
	in \secref{sec:discussion} and \secref{sec:summary}, respectively.

\section{Methods}    	
	\label{sec:method}
	We perform a suite of simulations of photoevaporating protoplanetary disks
	with various metallicities of $10^{-3}~\smetal \leq \metal \leq 10^{0.5}~\smetal $.
	We solve coupled equations of hydrodynamics, radiative transfer, and 
	non-equilibrium chemistry.
	
	We largely follow the method of Paper I, 
	except that we include X-ray radiation
	and add \ce{H2+} as a chemical species in the present study.
	In this section,
	we briefly summarize our model and
	refer the readers to Paper I for numerical methods.
	Details of the X-ray implementation are described in \appref{app:X-ray}.

	We take into account FUV, EUV, and X-ray irradiation from the central star.
	The central star is assumed to have 
	the stellar parameters tabulated in \tref{tab:model}.
        Although the stellar properties may well depend on metallicity,
        we adopt the fixed parameters in all our simulations in order to
        make it easy to compare the results directly.
	
	The FUV and EUV luminosities and the SED are the same as
	in Paper I. 
	We adopt the X-ray SED 
	presented in \cite{2007_Nomura_II}, which is derived by fitting the 
	observational {\it XMM-Newton} data 
	for TW Hydrae with using
	a two-temperature thin thermal plasma model \citep{1985_Mewe,1995_Liedahl}.
	In our model, 
	we set
	the minimum and maximum energy of the SED to be
	$\Emin = 0.1 \keV$ and $\Emax = 10 \keV$, respectively\footnote{We refer
          to photons with $h\nu \geq 0.1 \keV$ as X-rays in the present study.}.

	The disk gas is composed of the eight chemical species: 
	H, \ce{H+}, \ce{H2}, \ce{H2+}, CO, O, \ce{C+}, and electrons.
	Note that we add \ce{H2+} in order to follow \ce{H2} ionization by X-rays.
	Hereafter,
	we refer to 
	H, \ce{H+}, \ce{H2}, and \ce{H2+} 
	as H-bearing species
	and CO, O, and \ce{C+} as metal species.
	The amounts of the dust and heavy elements are 
	determined by the ISM values for our solar metallicity disk, 
	and assumed to be proportional to the ratio of
	the metallicity $\metal$
	to the local interstellar metallicity $\smetal$.
	Thus, 
	we use the dust to gas mass ratio  $\dgratio = 0.01 \times \metal/\smetal$,
	and the gas-phase elemental abundances of 
	carbon $ \abn{C} = 0.927\e{-4}~ \metal/\smetal$
	and oxygen $\abn{O} = 3.568\e{-4}~ \metal /\smetal$
	\citep{1994_Pollack,2000_Omukai}.
	The dust-to-gas mass ratio and the elemental abundances
	are the same as in Paper I.

	\begin{table}[htp]
		\caption{Properties of the model}
		\begin{center}
		\begin{tabular}{l  r}	\hline \hline
		{\bf Stellar parameters}	&						\\
		Stellar mass			&	
		$ 0.5~M_\odot$									\\
		Stellar radius			&	
		$ 2~R_\odot$									\\
		FUV luminosity			&	
		$  3\e{32}~\unit{erg}{}~\unit{s}{-1}$					\\
		EUV photon number rate			&	
		$  6\e{41}~\unit{s}{-1}$							\\	
		X-ray luminosity		&	
		$ 10^{30}~\unit{erg}{}~\unit{s}{-1}$\\
		\hline
		{\bf Gas/dust properties}	&						\\
		Species				&	 \ce{H}, \ce{H+}, \ce{H2}, \ce{H2+}, 
		CO, \ce{O}, \ce{C+}, \ce{e-}	\\
		Carbon abundance		&	
		$ 0.927\e{-4} \times Z/ Z_\odot$					\\
		Oxygen abundance		&	
		$ 3.568\e{-4} \times Z/Z_\odot$						\\	
		Dust to gas mass ratio	&	
		$ 0.01 \times Z/ Z_\odot$							\\
		\hline
		\end{tabular}
		\end{center}
		\label{tab:model}
	\end{table}

	The simulations are performed in
	2D spherical polar coordinates $(r, ~\theta)$.
	The disk is assumed to be symmetric 
	around the rotational axis $(\theta = 0)$
	and to the mid-plane $(\theta = \pi /2)$.
	The time evolution of the gas density, 
	velocity, energy, 
	and chemical abundances are solved.
	Although the computational domain is defined in 2D,
        we solve the azimuthal velocity evolution
	as well as the poloidal velocity $\bm{v}_\text{p} = (v_r,~ v_\theta)$.
	In the energy equation, 
	relevant heating/cooling sources are included (Paper I).
	For the chemical evolution, 
	we take into account both the advection and chemical reactions.

	X-ray heating, X-ray ionization,
	and the associated chemical reactions involving \ce{H2+}
	are added to our chemistry model.
	We describe the implementation 
	of these physical processes 
	in \appref{app:X-ray}.

	The equation of state is given as in Paper I,
	but the ratio of specific heat $\gamma$ is 
	calculated with considering the contribution of \ce{H2+},
	\begin{equation}
		\gamma = 1 + \frac{\abn{\ce{H}} + \abn{\ce{H+}} + \abn{\ce{H2}} + \abn{\ce{H2+}} + \abn{e}}
		{\frac{3}{2} \abn{\ce{H}} + \frac{3}{2} \abn{\ce{H+}} + \frac{5}{2}\abn{\ce{H2}} 
		+ \frac{5}{2}\abn{\ce{H2+}} + \frac{3}{2} \abn{e}}.	\label{eq:specific_heat}
	\end{equation}

	FUV, EUV, and X-ray radiative transfer is solved by ray-tracing but with neglecting scattering.
	The diffuse EUV due to recombination of hydrogen has been proposed 
	as an important component to drive photoevaporation
	\citep{1994_Hollenbach}, 
	but a recent study by \cite{2013_Tanaka} shows that 
	the direct EUV is dominant over the diffuse one. 
	\cite{1994_Hollenbach} uses a disk model which has an infinitesimally thin 
	disk structure in the region outer than the gravitational radius 
	and it has a finite scale height 
	in the region inner than the radius. 
	\cite{2013_Tanaka} uses another one where 
	disk scale height is finite at any distance. 
	The difference in the conclusions between these studies
	can be partly derived from the difference of the adopted disk models.
	This implies that 
	the direct EUV can be dominant in the outer region
	unless there is a geometrically thick region 
	which completely attenuates the direct EUV
	as in \cite{1994_Hollenbach}.
	We note that \cite{2017_Hollenbach} discusses 
	the causes for the differences in conclusions between these two studies.
	Our disk model does not have the geometrically thick region.
	Therefore, 
	we neglect the diffuse EUV component in our model.
	The FUV and EUV radiative transfer is done as in Paper I, whereas
	the X-ray radiation transfer is described  in \appref{app:X-ray}.
	The dust temperature is calculated by
	following radiative transfer for the direct stellar irradiation component 
	and the diffusive dust (re-)emission component, using a hybrid scheme of \cite{2010_Kuiper}.

	We set the computational domain on $r = [1, ~400] \AU$ 
	and $\theta = [0, ~\pi/2] {\rm ~ rad}$.
	Since the gravitational radius for a $0.5\, M_\odot$ central star is $\sim 0.7\AU$
	for $10^4\Kelvin$ ionized gas \citep{2003_Liffman},
	setting the inner boundary at $1\AU $
	may result in an underestimate of the 
	photoevaporative mass loss rate.
	However, 
	the contribution from within $ 1\AU$ is only a small fraction 
	of the total mass loss rate.\footnote{In Paper I,
	we ran simulations with smaller inner boundaries of $0.1\AU$, $0.35\AU$, and $0.5\AU$.
	We found that the density of the ionized atmosphere is too small to 
	shield EUV photons and the EUV photons actually reach the outer region.
	We thus justify our setting of the inner boundary at $1\AU$.}
	Therefore, we set the inner boundary at $r = 1 \AU$ in our simulations. 
	\cite{2017_Hollenbach} show that using a finite-size sink 
	would result in missing the attenuation of the direct EUV inside it.
	This effect may reduce the direct EUV reaching the outer disk 
	and yield a smaller EUV photoevaporation rate by a factor.
	Ideally, it would be better to define a computational domain which extends down to 
	the stellar surface, but it is clearly beyond the limitation of the currently available
        numerical methods.
        We do not set the strict surface boundary conditions but note here
        that the accreting gas to the stellar surface might shield 
	the high energy photons \citep{2018_Takasao}.         

	We choose a sufficiently large 
	domain to avoid spurious reflection of soundwaves and gas flows.
	In Paper I, we show that a sufficiently large outer boundary can eliminate the
        spurious effect.

	We run a set of simulations 
	where all of the photoionization heating by EUV (hereafter, EUV heating),
	photoelectric heating by FUV (hereafter, FUV heating),
	and X-ray heating are taken into account.
	In order to isolate the effects of 
	X-ray heating, 
	we also run simulations without FUV heating.
	The resulting photoevaporation rates 
	are compared with the results of Paper I, 
	where X-rays are not included.
	Hereafter, we label the sets of our simulations 
	according to which (or the combination) of FUV, EUV, and X-ray heating 
	is taken into account. A simulation is labeled as ``Run YYY'', 
	where ``YYY'' specifies which of the photo-heating sources are included.
	For example, in ``Run FEX'', 
	all the photo-heating effects are included
	If only EUV heating is taken into account, 
	the simulation is referred to as ``Run E''.
	In addition, when we refer to a simulation 
	with metallicity $\metal = 10^C \, \smetal$, 
	we append ``/Z\,$C$'' to the above labels.
	For instance, when we refer to 
	the simulation with $\metal = 10^{-0.5}\, \smetal$
	in which FUV and EUV heating are taken into account,
	we refer to the simulation as ``Run FE/Z\,-0.5''.

\section{RESULTS}			\label{sec:result}
	In this section, 
	we study the photoevaporation rate
	and the structure of PPDs.
        We first present the results of the solar metallicity case in \secref{sec:result1}.
        We then discuss the metallicity dependence in
	\secref{sec:result2} and \secref{sec:result3}.

	\subsection{Solar Metallicity Disks}	\label{sec:result1}
		\begin{figure}[htbp]
		\begin{center}
		\includegraphics[width=\linewidth, clip]{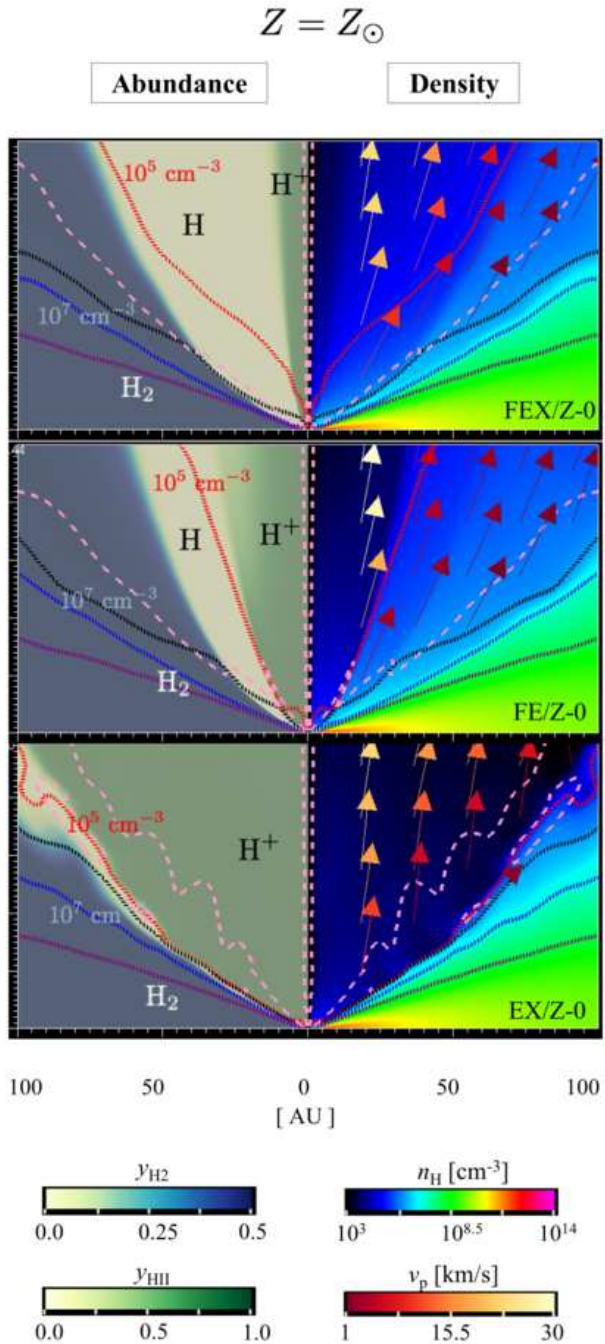}
		\caption{
			Solar-metallicity disk structures in Run FEX/Z0 (top),
			Run FE/Z0 (middle), and Run EX/Z0 (bottom).
			The left panels show the chemical structures
			and the right panels show the density structures.
			The arrows indicate the velocity fields with 
			the poloidal velocities of $\vp > 1 \kms$.
			The dotted lines are the density contours 
			of $\nh = 10^5\cm{-3}$ (red), $\nh = 10^6\cm{-3}$ (black),
			$\nh = 10^7\cm{-3}$ (blue), and $\nh = 10^8\cm{-3}$ (purple).
			The pink dashed lines indicate
			the sonic surface.
		}
		\label{fig:evaporations}
		\end{center}
		\end{figure}

		\fref{fig:evaporations}
		shows that
		photoevaporative flows are excited in all 
		the cases.
		Dense neutral photoevaporative flows
		composed of atomic and molecular hydrogen
		are driven in Run FEX/Z0 and Run FE/Z0,
		but not 
		in Run EX/Z0.
		Ionized photoevaporative flows,
		which consist of ionized hydrogen,
		are driven in all the three simulations.
		FUV heating is an important process
		to drive the neutral flows.
		X-ray heating is ineffective to drive photoevaporation
		in Run EX/Z0, where EUV-driven flows are excited only from
		the low density, disk surface regions. 
		
		Following Paper I,
		we measure the photoevaporation rate $\mdotph$ 
		by integrating the mass flux normal
		to a spherical surface $S$
		\begin{equation}
			\dot{M}_{\rm ph} = \int_{S} d\bm{S} \cdot \rho \bm{v}
			 =r_S^2 \int_{S} d\theta d\phi  \sin \theta \rho v_r  , \label{eq:pratesim}
		\end{equation}
		where $d\bm{S}$ is an integral element vector of the surface
		and $r_S$ is the radius of the sphere. 
		A gas parcel is regarded as unbound, and
		its contribution to $\mdotph$ is counted
		only if 
		the total specific enthalpy of the gas
		\begin{equation}
			\eta	= \frac{1}{2} \bm{v}^2 + \frac{\gamma}{\gamma -1 } c_s^2 - \frac{GM_*}{r}  ,
			\label{eq:eta}
		\end{equation}
		is positive at the surface 
		\citep{2003_Liffman}.
		With this condition,
		contributions from the bound disk region $(\eta < 0)$ to $\mdotph$
                are effectively excluded.

		In Section 3.3 of Paper I, 
		we show that photoevaporation rates measured as in \eqnref{eq:pratesim}
		generally increase with $r_S$.
		This is because the temperatures in the launching regions of photoevaporative flows,
		also called base temperatures,
		are generally larger than 
		necessary to escape from the gravity
		in the outer region beyond the gravitational radius;
		There are contributions to the mass loss in the outer region.
		Thus, photoevaporation rates should be estimated 
		with different $r_S$ to count the contributions.
		We calculate  $\mdotph$ 
		with setting $r_S = 20\AU,~100\AU, ~200\AU$
		for each of Run FEX, Run FE, and Run EX.
		For the solar metallicity disk, 
		we also perform Run X/Z0 
		to examine explicitly the contribution of X-ray-driven flows to $\mdotph$.
		The photoevaporation rates with $r_S= 20\AU$ include
		contribution from the inner part, 
		while those with $r_S= 100\AU$ and $r_S=200\AU$ include
		contribution from the whole region of a disk.
		Generally, in our simulations, 
		photoevaporation rates vary in time
		for the first $\sim 5000$ years, but then converge afterward.
		We take a time-averaged 
		photoevaporation rate over $5000$ years.
		The resulting photoevaporation rates  		
		for Run FEX/Z0, Run FE/Z0, Run EX/Z0, and Run X/Z0
		are given in \tref{tab:prate}
                for each case of $r_S = 20\AU$, $r_S = 100\AU$, and $r_S = 200\AU$,		
		and are also plotted in \fref{fig:prate_z}.
                
	\begin{table}[htp]
		\caption{Resulting photoevaporation rates $\mdotph$
		of the solar metallicity disk in $\myr$}
		\begin{center}
		\begin{tabular}{c | c c c c}
			$r_S$	&	Run FEX/Z0	&	Run FE/Z0	&	Run EX/Z0  
			& Run X/Z0
			\\ \hline\hline
		$20\AU$  	& $6\e{-9}$	&	$3\e{-9}$	&	$2\e{-10}$ &$5\e{-12}$\\
		$100\AU$ & $2 \e{-8}$	& 	$2\e{-8}$	&	$1\e{-9} $ & $1\e{-11}$\\
		$200\AU$ &$3 \e{-8}$	&	$3\e{-8}$	& 	$2\e{-9}$  &$2\e{-11}$\\ \hline		
		\end{tabular}
		\end{center}
		\label{tab:prate}
	\end{table}
        
		We note that  $\mdotph$ in Run FEX/Z0 
		is slightly higher than in Run FE/Z0. 
		Also,  $\mdotph$ in Run EX/Z0
		is about an order of magnitude smaller
		than in Run FEX/Z0 and in Run FE/Z0.
		We find a very small $\mdotph$ in Run X/Z0.
		Overall, these results suggest that 
		FUV is a crucial radiation component to produce a high $\mdotph$,
		whereas X-rays give a minor contribution to $\mdotph$, although 
		X-rays affect the structure of the neutral flows
		(see Run FEX/Z0 and Run FE/Z0 in \fref{fig:evaporations}).

		\begin{figure*}[htbp]
		\begin{center}
		\includegraphics[clip, width = \linewidth]{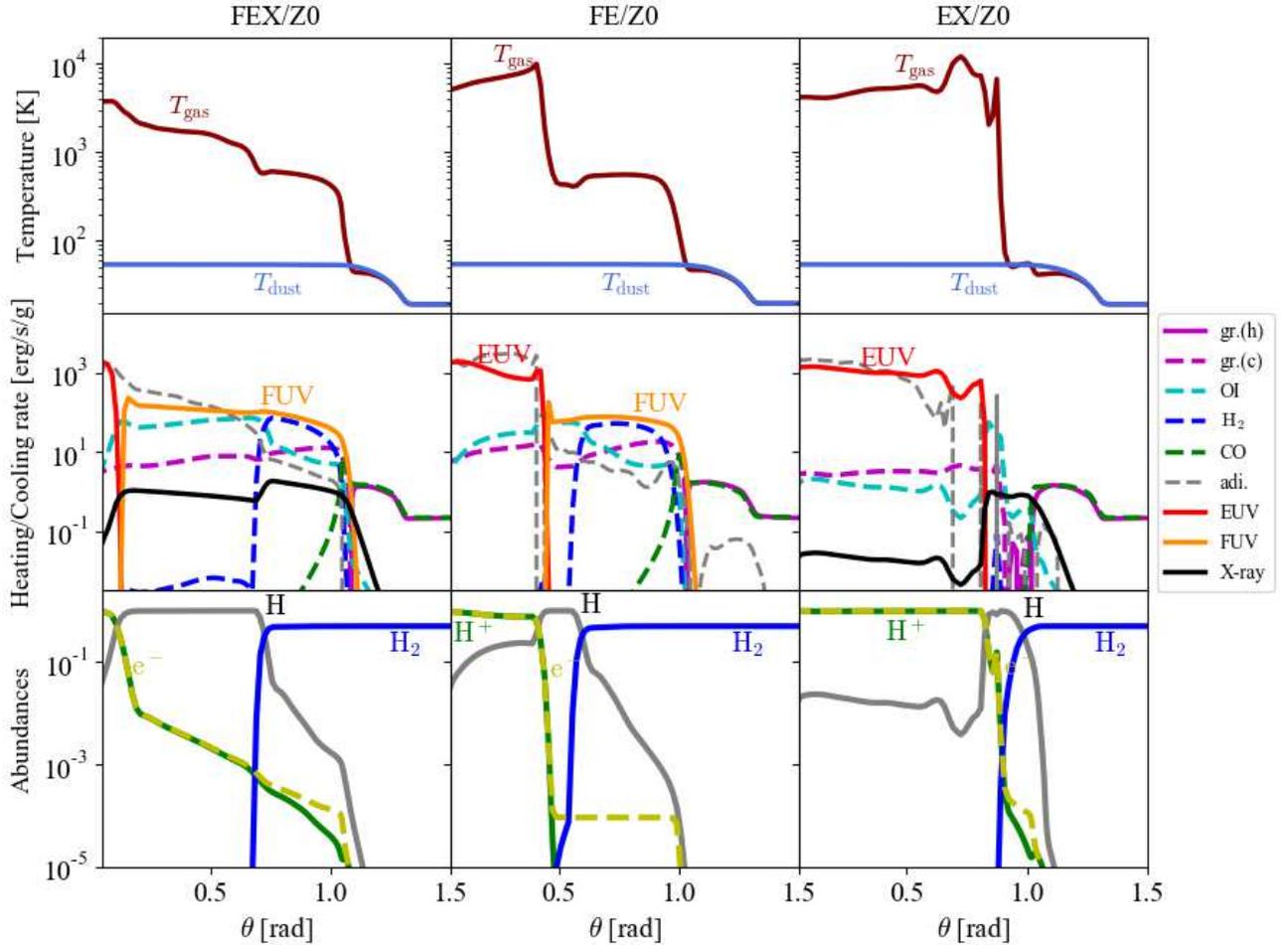}
		\caption{Meridional distributions of the physical quantities at $r = 100\AU$: 
		gas and dust temperatures (top), specific heating/cooling rates (middle),
		and chemical abundances of H, \ce{H+}, \ce{H2}, and electrons (bottom).
		In the middle row, 
		the solid lines show
		the heating rates of EUV (red), FUV (orange), X-ray (black), and dust-gas collision (purple), 
		while the dashed lines show
		the cooling rates of dust-gas collision (purple), OI (cyan), \ce{H2} (blue),
		CO (green), and adiabatic cooling (gray). 
		Each column shows 
		the physical quantities
		of Run FEX/Z0, Run FE/Z0, and Run EX/Z0
		for solar metallicity disks
		from the left to the right.
		}
		\label{fig:theatcol}
		\end{center}
		\end{figure*}

                The X-ray heating rate in Run EX/Z0 is smaller
		than the FUV heating rate in Run FEX/Z0 and in Run FE/Z0
		(see the second row of \fref{fig:theatcol}).
		The neutral regions 
		in the disk are not heated by X-rays
		to sufficiently high temperatures
		to escape from the star-disk system.
		The gas temperatures are 
		nearly coupled with the dust temperatures 
		in this region (the top right panel of \fref{fig:theatcol}).
		We conclude that 
		X-ray heating itself is not efficient 
		to excite photoevaporative flows.
		Only the EUV-driven, ionized flows contribute to the mass loss
		in Run EX/Z0.		
		\fref{fig:evaporations} shows that
                the ionized flows have densities
		typically several orders of magnitude 
		smaller than the neutral flows.
		The resulting $\mdotph$ of Run EX/Z0 is
		much smaller than those of Run FEX/Z0 or Run FE/Z0,
		where 
	        FUV-driven neutral flows give a large contribution to $\mdotph$.

		FUV heating is dominant in the neutral regions
		in Run FEX/Z0 and
		Run FE/Z0 (\fref{fig:theatcol}).
		By studying these runs in detail, we find
                that the FUV heating rate is higher in Run FEX/Z0 than in Run FE/Z0 
		especially in the regions close to the ionization front.
		There, the gas is weakly ionized by X-rays.
		With the electron abundance slightly increased,
		dust grains recombine,
		the positive charges of dust grains are reduced,
		and then the photoelectric effect efficiency is increased
                \citep{2009_GortiHollenbach}.
                
		To summarize,
		X-ray ionization effectively strengthens  
		FUV heating in neutral regions in a disk
		by increasing the photoelectric effect efficiency.
		Owing to the strengthened FUV heating,
		the temperatures of the neutral regions
		are higher in Run FEX/Z0
		than in Run FE/Z0.
                With the combined FUV+X-ray heating effect,
                the neutral gas near the central star
                evaporates in Run FEX/Z0.
		We have checked and found that there is a large difference  
		in $\mdotph$ measured with $r_S  = 20\AU$
                between Run FEX/Z0 and Run FE/Z0 (\tref{tab:prate}).
		However, in these runs, the photoevaporative flows excited in the inner regions
                accounts for only a small fraction of the total mass loss rate,
                and a dominant contribution comes from outer disk regions.
		Therefore, there is a small
		difference in $\mdotph$ between Run FEX/Z0 and Run FE/Z0, 
		when measured with $r_S = 100\AU$ or $r_S = 200\AU$.

	\subsection{FUV Heating in Low Metallicity Disks}
	\label{sec:result2}
	\begin{figure}[htbp]
		\centering
		\includegraphics[clip, width = \linewidth]{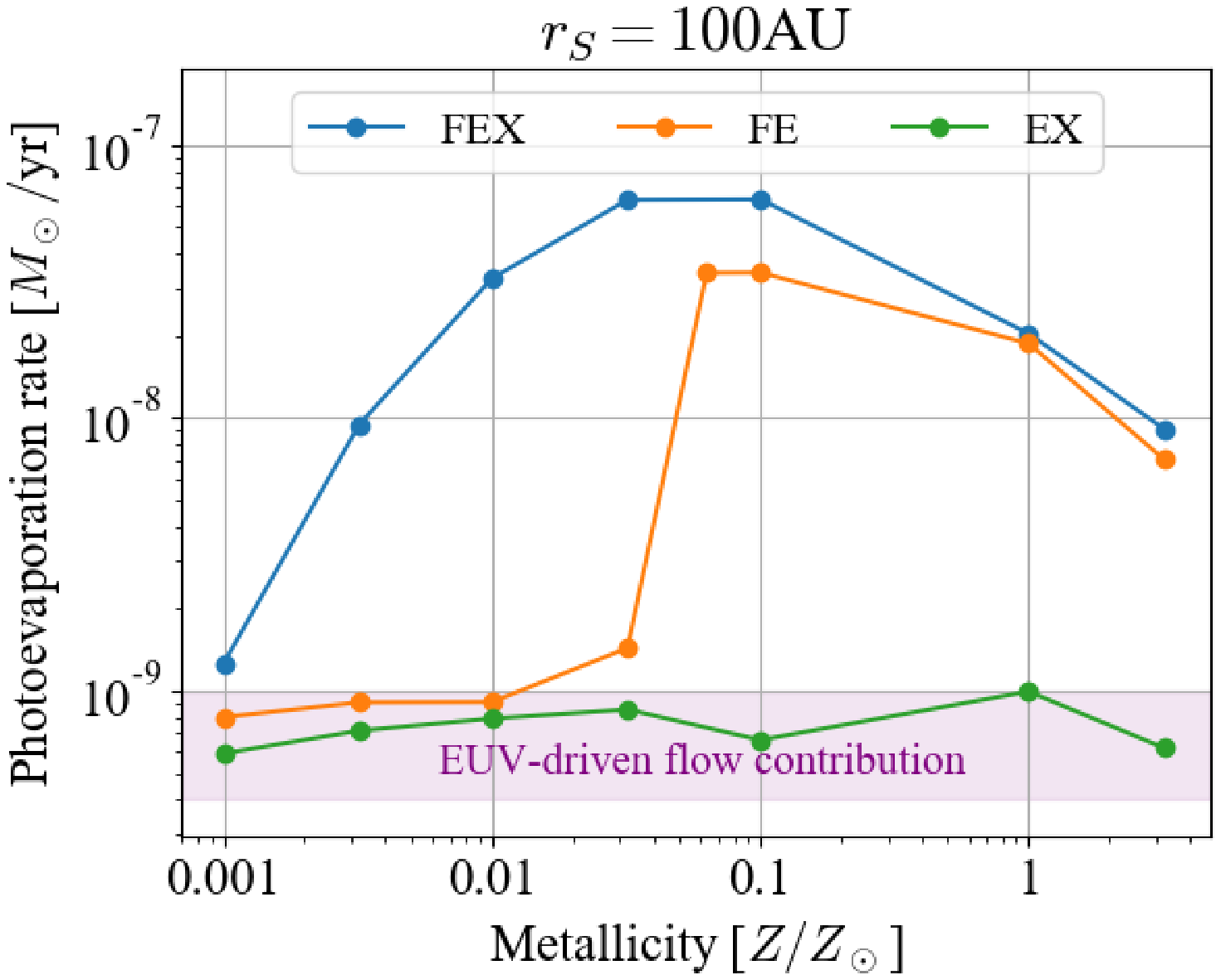}
\includegraphics[clip, width = \linewidth]{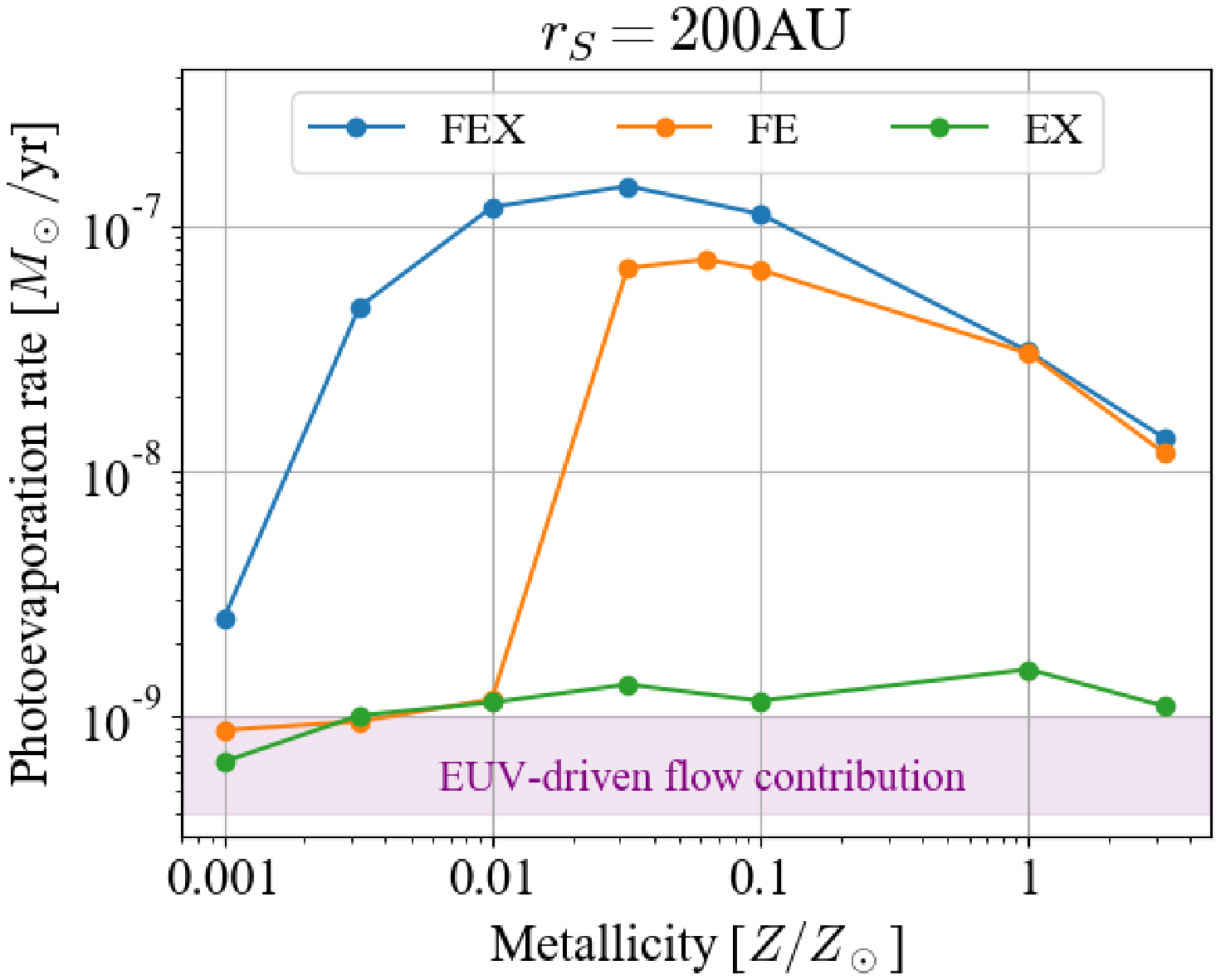}
		\caption{
		Each panel shows
		the difference in the metallicity dependences of the photoevaporation rates 
		for Run FEX (blue), Run FE (orange), and EX (green).
		The purple regions show approximate 
		EUV-driven flow contributions ($\dot{M}_\text{EUV} \simeq 0.4 - 1 \e{-9} \myr$),
		separating the EUV photoevaporation rates 
		from the total photoevaporation rates.
		The photoevaporation rates  
		are calculated with $r_S = 100\AU$ and $r_S = 200\AU$
		in the top and bottom panels, respectively.
		}
		\label{fig:prate_z}
	\end{figure}

                FUV radiation can reach the deeper interior of a disk with a lower metallicity 
		because the amount of dust grains and hence its opacity are correspondingly smaller.
		This results in exciting photoevaporative flows with higher densities.
		On the other hand, the FUV heating becomes progressively inefficient at low metallicities
		compared to dust-gas collisional cooling.
		The relatively inefficient FUV heating lowers the base temperatures, and 
		FUV-driven flows are not excited in the ``cool'' disk.
		Consequently, $\mdotph$ of Run FE increases  
		in the range of $10^{-1}\, \smetal \lesssim \metal \lesssim 10^{0.5}\, \smetal $
		but decreases in the range of $10^{-2}\, \smetal \lesssim \metal \lesssim 10^{-1}\, \smetal $		
		(see \fref{fig:prate_z}).
		Note that a major contribution to the mass loss rate comes from FUV-driven photoevaporative flows 
		in runs with $\metal \gtrsim 10^{-2} ~\smetal$.
		EUV-driven flows are important in runs 
                with $\metal \lesssim 10^{-2} ~\smetal$, where the metallicity dependence is small.
		
		Although including X-rays affects
		the metallicity dependence of $\mdotph$ 
		(compare Run FEX and Run FE in \fref{fig:prate_z}),
		the overall trend appears quite similar;
		$\mdotph$ of Run FEX increases 
		as metallicity decreases in $\metal \gtrsim 10^{-1.5}\, \smetal$
		because of the reduced opacity effect,
		while $\mdotph$ decreases with metallicity in $\metal \lesssim 10^{-1.5}\, \smetal$
		because FUV photoelectric heating becomes
		less effective than dust-gas collisional cooling.

                \subsection{The Effect of X-ray Ionization}
                \label{sec:result3}
                There is a significant difference between Run FEX and Run FE
                at very low metallicities of $\metal \lesssim 10^{-1.5}\, \smetal$.
		In Run FEX, X-ray ionization raises
		the electron abundance in the neutral region of the disk,
		and 
		effectively {\it strengthens} the FUV heating.
                The gas temperature in the neutral region is higher
		than in Run FE.
		This causes the neutral gas to 
		evaporate from even 
		closer regions to the star.
		However, with the low metallicities of $\metal \gtrsim 10^{-1.5}\, \smetal$,
		the contribution from the inner region is still 
		small with respect to the total photoevaporation rate (\fref{fig:prate_z}).
		This is similar to the case with solar metallicity as discussed in \secref{sec:result1}.
		In other words,
		FUV heating can drive neutral photoevaporative flows 
		even without the strengthening effect of X-rays
		in this metallicity range;
		the photoevaporation rates of Run FEX and Run FE
		are close to each other. 
		This is in good contrast to the runs with
		$ 10^{-2.5}\,\smetal \lesssim \metal \lesssim 10^{-1.5}\,\smetal$.

\begin{figure}[htbp]
		\begin{center}
			\includegraphics[clip, width = \linewidth]{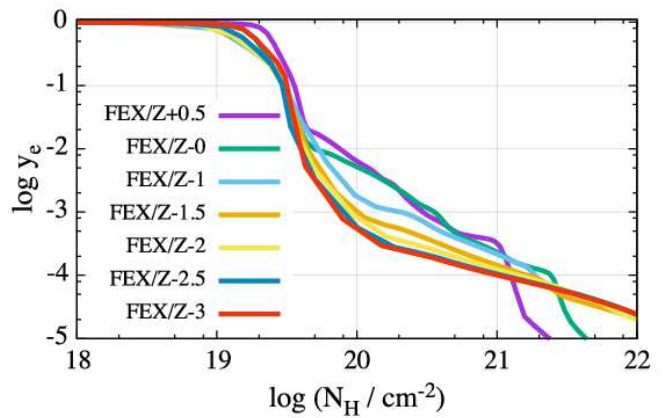}
		\caption{Meridional distributions of electron abundance at $r = 100\AU$
		for various metallicity disks.
		All of the data are taken from Run FEX.
		Note that we here use
		hydrogen nuclei column density $\col{H}$ 
		instead of $\theta$ unlike in \fref{fig:theatcol}. 
		}
		\label{fig:eabun}
		\end{center}
		\end{figure}

		In the range of $\metal \lesssim 10^{-1.5} \, \smetal$, 
		$\mdotph$ of Run FEX
		decreases with metallicity because of the efficient dust-gas collisional cooling 
		relative to FUV heating, 
		but we find a more gradual decline of $\mdotph$ than in Run FE.
		In Run FEX, 
		the electron abundance in the neutral region is increased 
		by the effect of X-rays.
		Hydrogen is the dominant X-ray absorber,
		and thus
		the electron abundance in the neutral region
		is essentially independent of metallicity 
		(\fref{fig:eabun}).
		Since the photoelectric effect efficiency 
		depends on electron density only through 
		the ratio of
		the dust/PAH photoionization rate to the dust/PAH recombination rate 
		$\gamma_\text{pe} = G_\text{FUV} \sqrt{T} / \nspe{e}$
		(see Eq.[41] of Paper I for details),
		it is not affected by 
		metallicity, at least explicitly, when the electron abundance
                is largely set by hydrogen ionization.
		As a result,
		FUV heating remains effective even at low metallicities
                ($ \metal \lesssim 10^{-1.5}\,\smetal$)
		in Run FEX than in Run FE, 
		and thus
		$\mdotph$ in Run FEX
		decreases more gradually.

		We find a large difference in $\mdotph$ 
		between Run FEX and Run FE, 
		especially in the range $ 10^{-2.5}\,\smetal \lesssim \metal \lesssim 10^{-1.5}\,\smetal$.
		This can be attributed 
		to the effect of the X-ray radiation
		through partial ionization.
		For example,
		in the disk with $\metal = 10^{-2} \, \smetal$,
		the electron abundance in Run FEX/Z-2 
		is about two orders of magnitude larger 
                than in Run FE/Z-2 
		in the low density part ($\nh \sim 10^5 - 10^6 \cm{-3}$)
		at around $100\AU$.
		Correspondingly, 
		the ratio of the dust photoionization rate to the dust 
		recombination rate $\gamma_\text{pe} = G_\text{FUV} \sqrt{T}/n_\text{e}$ 
		(cf. Appendix A of Paper I)
		is small, 
                with the typical value of $\sim 10^3 (n_\text{e} / 100\cm{-3})^{-1}$ in the low-density region.
                This is about two orders of magnitude smaller
		than in Run FE/Z-2.
		Therefore, 
		the photoelectric effect efficiency 
		\begin{eqnarray}
			   \epsilon_{\rm pe} &&
			   =  \left[ \frac{4.87\e{-2}}{1+4\e{-3} ~ \gamma_{\rm pe}^{~0.73}} \right.  \nonumber \\
			&&+ \left. \frac{3.65\e{-2}(T/10^4~{\rm K})^{0.7}}{1+2\e{-4} ~ \gamma_{\rm pe}} \right]  ,
		\end{eqnarray}		
		is larger by about an order of magnitude \citep{1994_BakesTielens}.
                The temperature is increased
		by a factor of a few, and then
		the gas satisfies the enthalpy condition $\eta > 0$.
		\begin{figure*}[htbp]
		\begin{center}
		\includegraphics[clip, width = \linewidth]{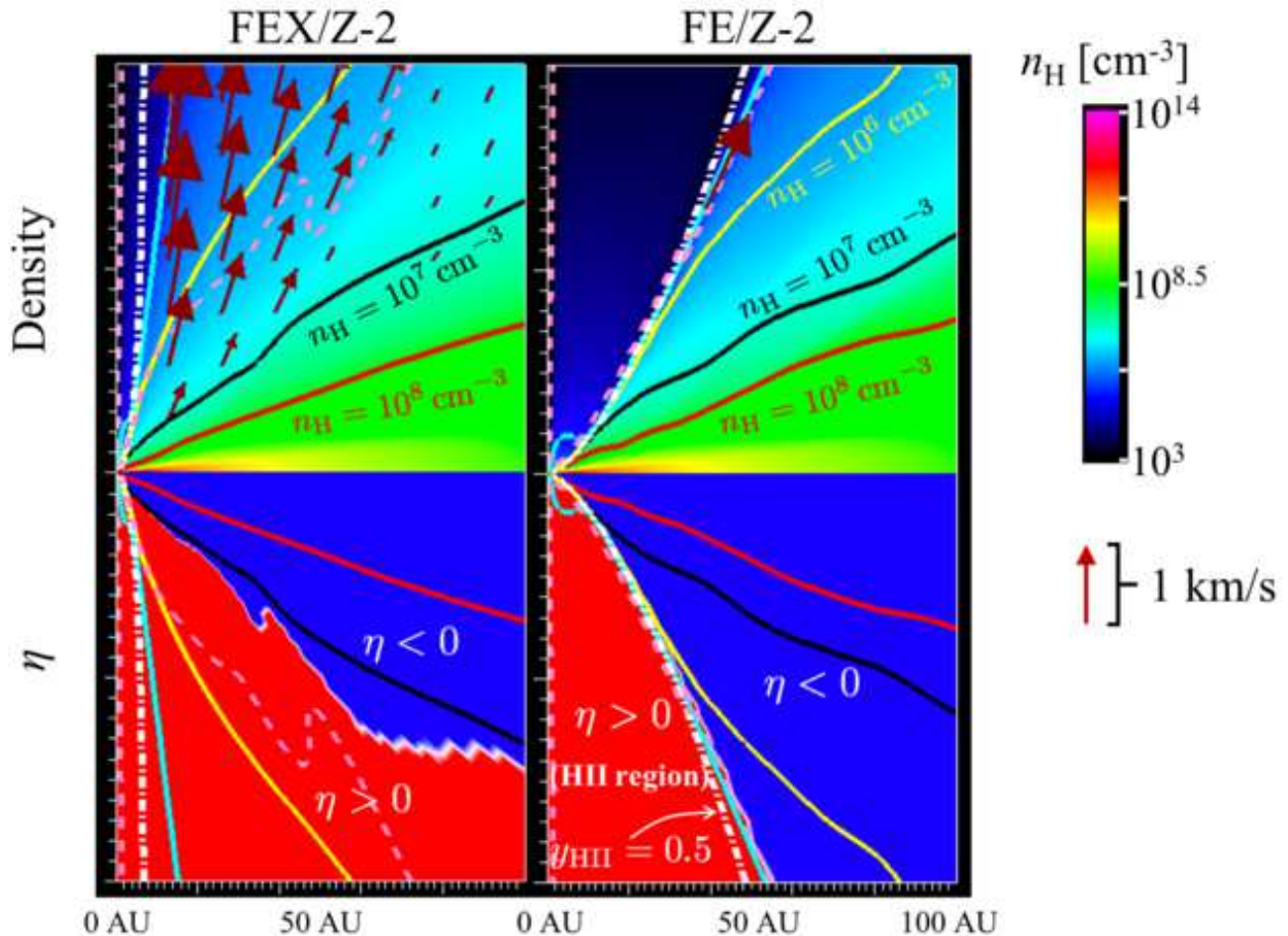}
		\caption{Snapshots of FEX/Z-2 (left) and FE/Z-2 (right). 
		The top panels show the density distributions, and 
		the bottom panels show the distributions of $\eta$ (\eqnref{eq:eta})
		where the red and blue regions indicate $\eta > 0$ (unbound) regions 
		and $\eta < 0$ (bound) regions, 
		respectively.
		The arrows show the velocity with $ 0.25\, \kms < \vp < 0.5 \, \kms$ and 
		they are scaled by its magnitude.
		The velocity field in the \HII~regions $(\abn{\HIImath} > 0.5)$ are not shown for clarity.
		The solid lines show density contours 
		of $\nh = 10^5 \cm{-3}$ (cyan), $\nh = 10^6 \cm{-3}$ (yellow),
		$\nh = 10^7 \cm{-3}$ (black), and $\nh = 10^8 \cm{-3}$ (red).
		Note that the velocity arrows and density contours are 
		drawn in different manners from \fref{fig:evaporations}.
		The white dot-dashed lines show contours of
		$\abn{\HIImath} = 0.5$, which is a rough boundary between 
		the \HII~region and the other.
		The pink dashed lines indicate the sonic surface.
		}
		\label{fig:eta}
		\end{center}
		\end{figure*}
		X-rays also affect
		other regions of the disk in a similar manner; 
		the total specific enthalpy of the gas is increased to be positive. 
		\fref{fig:eta} shows
		that the neutral region of Run FEX/Z-2 partly satisfies 
		$\eta > 0$, whereas
		the region with $\eta > 0 $ appears to overlap with
		the \HII~region in Run FE/Z-2.
		In runs with $\metal  = 10^{-2} \, \smetal $,
		incorporating X-ray ionization results in driving
		neutral photoevaporative flows, 
		which significantly contribute to the mass loss rate.
		Without X-rays, however, the neutral gas flows are not excited, and only
		EUV-driven ionized gas flows contribute to the mass loss.
		Consequently,
		Run FEX/Z-2 shows
		a significantly larger $\mdotph$ than 
		Run FE/Z-2.
                The same conclusion holds for
		Run FEX and Run FE with metallicities in the range
		of $ 10^{-2.5}\,\smetal \lesssim \metal \lesssim 10^{-1.5}\,\smetal$.

		In the very low metallicity of $\metal \lesssim 10^{-3}\, \smetal$,
		even though FUV heating is strengthened by the X-ray ionization effect,
		neutral flows are not excited
		because dust-gas collisional cooling becomes 
		more efficient than FUV heating.
		Therefore,
		there is not a significant difference between 
		the photoevaporation rates of 
		Run FEX/Z-3 and Run FE/Z-3.

	\subsection{PPD Lifetime}
	Regarding metallicity dependence 
	of PPD lifetimes,
	it is suggested that 
	typical lifetimes of protoplanetary disks 
	are $3\megayr$ 
	for solar metallicity disks and $1 \megayr$ for those with $\metal = 0.2 \,\smetal$
	\citep{2009_Yasui, 2010_Yasui, 2016_Yasui_I, 2016_Yasui_II}.
	This metallicity dependence of the lifetimes can be fit
	as $T_\text{life} \propto \metal ^ {0.7}$.
	In the present study, 
	the resulting photoevaporation rate
	of Run FEX
	has metallicity dependences of  
	$\mdotph \propto \metal ^{-0.6}$
	for 
	$r_S = 200\AU$, 
	in $0.1\, \smetal \leq \metal \leq 10^{0.5}\, \smetal$,
	while in Run FE  
	$\mdotph \propto \metal ^ {-0.4}$. 
	These metallicity dependences 
	are consistent with the observational metallicity dependence of the lifetimes
	because disk lifetimes are approximately calculated as 
	$T_\text{life} \propto \mdotph^{-2/3}$ \citep{2010_ErcolanoClarke}.
		Since X-ray radiation itself does not excite
		photoevaporative flows in a direct manner,
		the photoevaporation rate of Run EX 
		is largely contributed by the EUV-driven flows.
		Therefore, $\mdotph$ is generally
		metallicity-independent and is significantly smaller than 
		in Run FEX or Run FE, where
		FUV-driven flows contribute to the mass loss.
	This suggests that 
	in the case EUV heating mainly contributes to photoevaporation, 
	EUV and X-ray radiation does not cause 
	metallicity dependence in PPD lifetimes.
	Hence, if the metallicity dependence of the lifetimes 
	is originated from the metallicity dependence of photoevaporation,
	our model indicates that
	FUV photoevaporation has a major effect on the disk lifetimes.

\section{DISCUSSION}	\label{sec:discussion}

		Our conclusion regarding the effects of X-ray radiation
		is qualitatively consistent with 
		that of \cite{2009_GortiHollenbach} (hereafter, GH09),
		who conclude that
		X-ray photoionization increases 
		the efficiency of photoelectric heating
		and enhance the FUV photoevaporation rate.
                Although X-ray heating has been proposed as an important cause of photoevaporation
                in several studies
		\citep{2008_Ercolano,2009_Ercolano,2012_Owen},
                our direct comparison shows that 
		X-rays alone do not drive strong photoevaporation,
		in agreement with the conclusions of \cite{2004_Alexander}, 
		\cite{2009_GortiHollenbach}, and \cite{2017_Wang}.
	        In the following, we discuss the effect of a few elements associated
                with our X-ray radiation model.

	\subsection{X-Ray Spectral Hardness}
	\label{sec:speff}

                The hardness of the adopted X-ray spectrum 
		affects the strength of X-ray photoevaporation.
		Table 1 of \cite{2009_Ercolano}
		shows that the photoevaporation rate decreases 
		with the ``pre-screening'' column, 
		i.e., the hardness of the incident flux on a disk.
		With the pre-screening column of $\col{H} \geq 10^{21} \cm{-2}$,
		there are virtually no photons with $\lesssim 0.1\keV$ 
		reaching the interior of a disk 
		(see Figure 3 of Ercolano et al. 2009).
		In this case,
		the resulting photoevaporation rate is of the order of $10^{-11} \myr$,
		and is smaller by two orders of magnitude than 
		the case with the pre-screening column of 
		$\col{H} \leq 10^{20} \cm{-2}$,
		where the EUV component ($\leq 0.1 \keV$)
		also heats the disk.
		The result suggests that 
		using a hard X-ray spectrum 
		results in {\it inefficient} X-ray photoevaporation,
		as has been also pointed out by \cite{2015_Gorti}.
		The X-ray spectrum used in the present study is similar to that 
		with pre-screening column of $\col{H} \sim 10^{21} \cm{-2}$
		with which the X-ray photoevaporation rate is small $(\sim 10^{-11} \myr)$
		\citep{2009_Ercolano}.

	In order to examine 
	whether using a softer spectrum
	affects the resulting X-ray photoevaporation rate, 
	we additionally perform test simulations where
	our fiducial X-ray spectrum is shifted 
	to lower energies.  
	We shift the fiducial X-ray spectrum $F(E)$ as 
	$F(E\times \sqrt{10})$ and $F(E\times 10)$
	while fixing the total luminosity of $10^{30}\,\unit{erg}{}\,\unit{s}{-1}$.
	\begin{figure}[htbp]
	\begin{center}
	\includegraphics[clip, width = \linewidth]{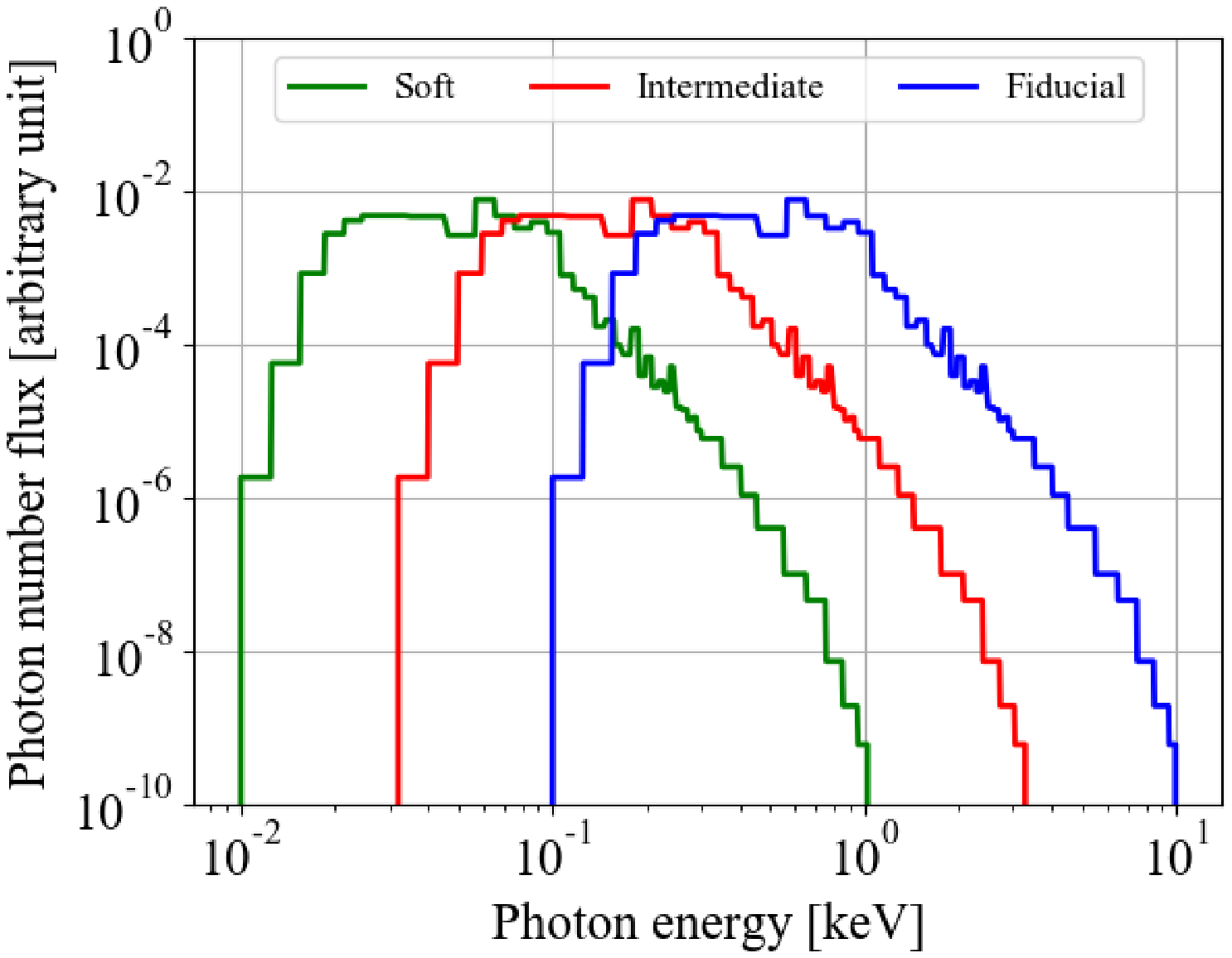}
	\caption{Fiducial SED (blue) and
	logarithmically shifted SEDs (red and green) for X-rays. 
	The shifted SEDs are given as 
	$F(E\times \sqrt{10})$ (red) and $F(E\times 10)$ (green), 
	where $F(E)$ is the fiducial SED function.
	}
	\label{fig:SEDs}
	\end{center}
	\end{figure}
	The shifted spectra are shown by
	the red and green lines in \fref{fig:SEDs}.
	Hereafter, we refer to the shifted spectrum colored in green 
	as the soft spectrum,
	and the other colored in red as the intermediate spectrum.
	The photo-heating rates are calculated by \eqnref{eq:X-rayheat}
	as in our fiducial model.
	For the heating efficiency,
	we use the same $f_h$ given by \eqnref{eq:fh}.
	Although $f_h$ might be larger for the softer spectra
	because all the primary electron energy goes into heat 
	through Coulomb interactions with the ambient electrons
	when $\abn{e} \sim 1$
	\citep{1996_Maloney}, 
	we do not model the heating efficiency as a function of photon energy.
	Instead, in \secref{sec:f_h},
	we study simulations with $f_h = 1$,
	corresponding to the limiting case where all the absorbed energy goes into heating.
	FUV heating is not taken into account 
	in our test simulations presented here.

	\begin{figure*}[htbp]
	\begin{center}
	\includegraphics[clip, width = \linewidth ]{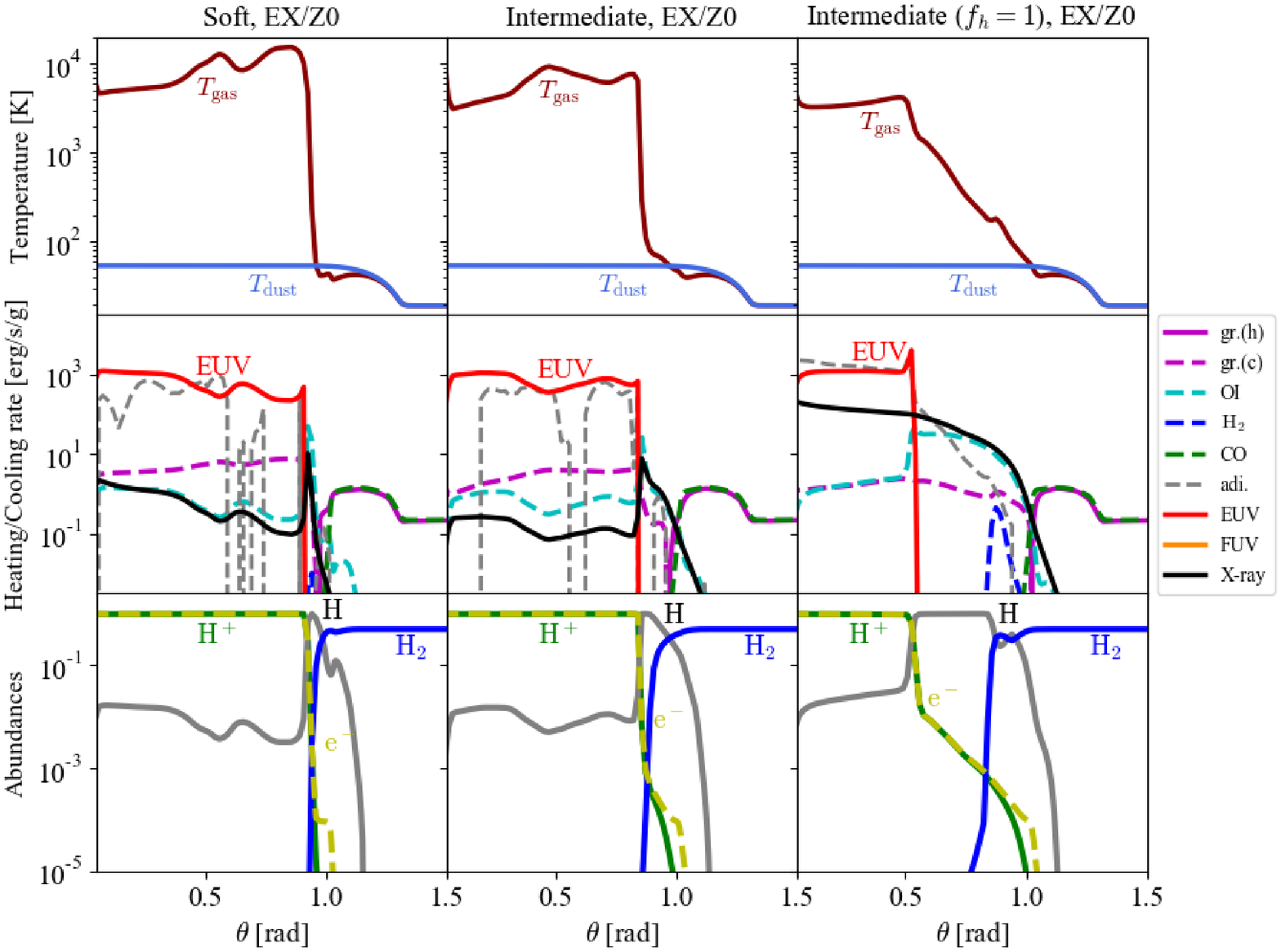}
	\caption{
	Meridional distributions of the physical quantities at $r = 100\AU$ in the simulations,  
	where the soft spectrum (left), intermediate spectrum (middle), 
	and intermediate spectrum with using $f_h = 1$ (right) 
	are used. 
	The panels are shown in the same manner as \fref{fig:theatcol}.
	Note that though the soft and intermediate spectra technically contain 
	the EUV component ($13.6\eV \leq h\nu \leq 100\eV$),
	we refer to the photo-heating 
	calculated with using the spectra
	as X-ray heating.}
	\label{fig:X-raytest}
	\end{center}
	\end{figure*}
	\fref{fig:X-raytest} shows that
	the maximum values of the photo-heating rates (the black lines)
	are larger by about an order of magnitude
	for the soft and intermediate spectra
	than that for our fiducial spectrum which is shown 
	in the right column of \fref{fig:theatcol}.
	The specific photo-heating rate is smaller for higher energy photons.
	\fref{fig:X-raytest} also shows
	that 
	low energy photons are nearly completely absorbed in the region 
	close to the ionization front.
        Since the cross section of the disk medium is larger
	for lower energy photons,
	adopting a softer spectrum results in
	a higher specific photo-heating rate, 
	but the low energy photons are absorbed in regions with small gas densities.

\begin{figure}[htbp]
		\begin{center}\includegraphics[clip, width = \linewidth]{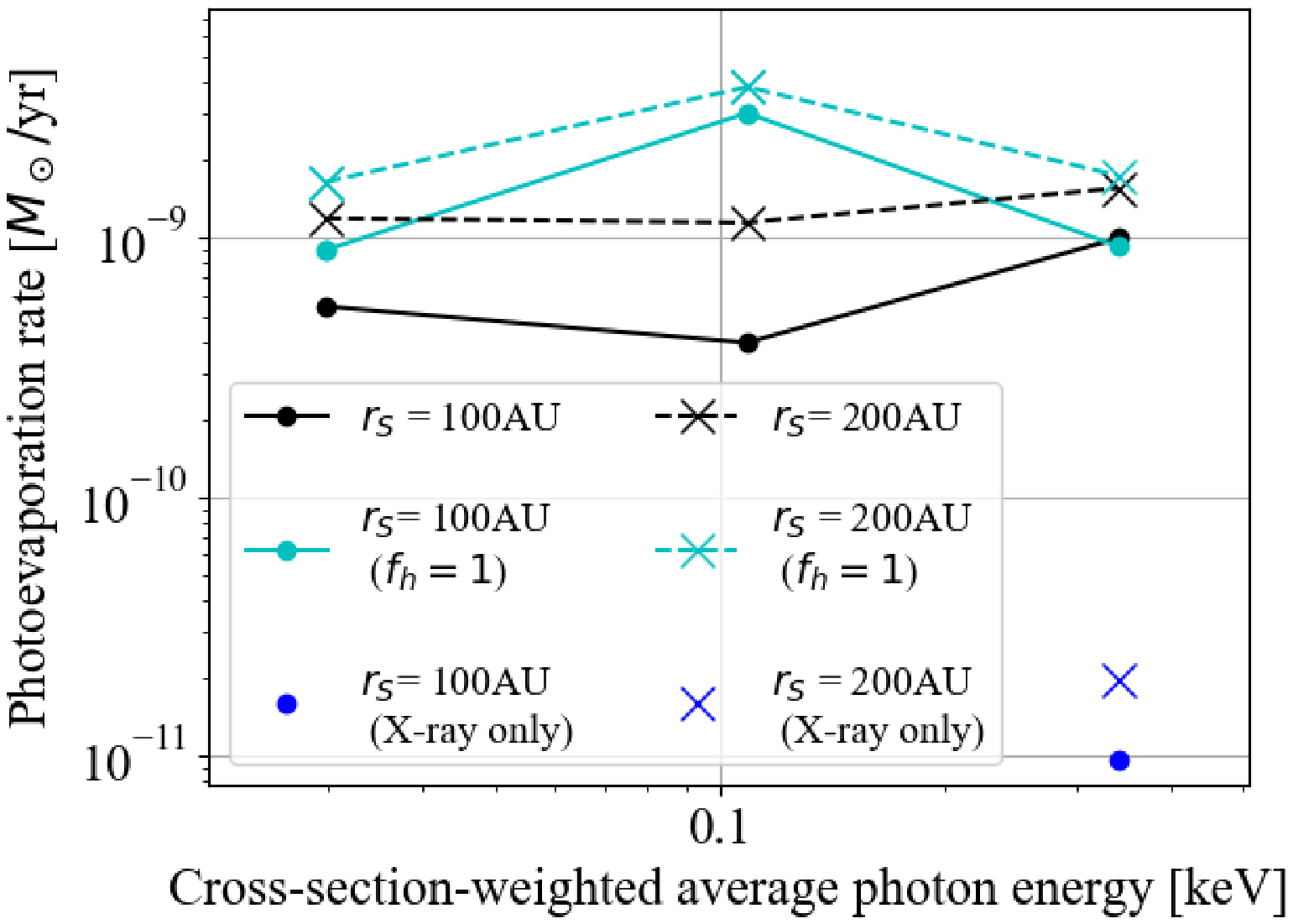}
		\caption{Resulting photoevaporation rates due to EUV and X-ray as functions of 
		the cross-section-weighted average photon energy
		of the soft, intermediate, and fiducial spectra.
		The black and cyan lines show the photoevaporation 
		rates derived from the simulations where 
		$f_h$ is calculated by \eqnref{eq:fh} and $f_h = 1$, respectively. 
		The blue points show the photoevaporation rates of 
		Run X/Z0 (\tref{tab:prate}).
		}
	\label{fig:pratediscussion}
	\end{center}
	\end{figure}

	The photo-heating raises the gas temperature 
	only in the region close to the ionization front,
	whereas a large part of the neutral region
	remains at relatively low temperatures, 
	as seen in \fref{fig:X-raytest}.
	The gas is not hot enough to launch
	neutral outflows, and 
	the EUV-driven,
	ionized flows dominantly contribute to mass loss rate.
	\fref{fig:pratediscussion} compares the photoevaporation rates 
	for the soft, intermediate, and fiducial spectra.
	Clearly, in our test simulations, 
	the spectral hardness does not critically affect the
        photoevaporation rate, although it changes the thermal and chemical structure of the disk
        (Fig. 7).

	\subsection{Effect of the Heating Efficiency $f_h$}
	\label{sec:f_h} 
	Another important factor is the heating efficiency $f_h$.
	Adopting $f_h = 1$ raises the photo-heating rate.
	We see in \fref{fig:X-raytest} that
	the neutral region has a higher temperature
	when using $f_h = 1$ (the right column)
	than when using $f_h$ of \eqnref{eq:fh} (the left and middle columns).
	Consequently,
	the photoevaporation rate with the intermediate spectrum with $f_h = 1$ are the highest
	(\fref{fig:pratediscussion}).
	Note that the cross-section-weighted mean energy $\bar{E}$ is
        $0.03\keV$ and $0.11\keV$ for 
	the soft and intermediate spectra, respectively. 
        For our fiducial spectrum, $\bar{E} = 0.34\keV$,
        $f_h = 1$ does not significantly change the photoevaporation rate.
	We conclude that high energy photons with $\gtrsim 0.1 \keV$ 
	are ineffective to heat the neutral gas and do not excite dense photoevaporative flows.
	The same is true for the soft and intermediate spectra.
	Our test simulations show that 
	the effective component for photoevaporative mass loss is not X-ray $(0.1\keV \leq h \nu \leq 10\keV)$,
	but EUV $(13.6\eV \leq h \nu \leq 100\eV)$.
        Hard EUV photons with $\sim 10^2\eV$ most efficiently drive disk mass loss.
	
	\fref{fig:pratediscussion} compares directly $\mdotph$ in the above test runs
        with that of Run X/Z0. Again, we confirm that X-rays are ineffective to excite photoevaporative flows. 
        When FUV heating is absent (a somewhat artificial condition, but for the purpose of direct comparison), 
        EUV mainly contributes to the photoevaporative mass loss. 
	This is consistent with 
	Table 1 of \cite{2009_Ercolano}, where we find
	the photoevaporation rates are significantly low when
	the EUV component is screened out (see also \cite{2015_Gorti}).

	In \cite{2009_Ercolano},
	the EUV component reaches the disk surface 
	if the pre-screening column density 
	$\col{H}$ is smaller than $10^{20} \cm{-2} $.
	The spectrum model of \cite{2009_Ercolano} with $\col{H} = 10^{19} -10^{20} \cm{-2}$
	is similar to our intermediate spectrum. 
	Their resulting photoevaporation rate of \cite{2009_Ercolano} 
	is $\mdotph \simeq 4 \e{-9} \myr$, 
	which is close to 
	$\mdotph$ of the test simulation with the intermediate spectrum and $f_h = 1$ (\fref{fig:pratediscussion}).
	Further, the photoevaporation rate differs only by a factor of two
	from that of \cite{2012_Owen},
	where an ``ionization parameter'' approach is used
	with the unscreened spectrum of \cite{2009_Ercolano} to calculate gas temperatures
	in the hydrodynamics simulations. 
	The agreement between the photoevaporation rates in \cite{2012_Owen} and the present study
	implies that,
	for a spectrum with a large amount of $\sim 0.1\keV$ photons,
	the ionization parameter approach 
	yields essentially the same results 
	as those derived with 
	heating and radiative cooling in a consistent manner. 
	Nonetheless, 
	self-consistent calculations such as ours are necessary 
	to investigate the relative importance between X-ray and FUV photons.

	The opening angle of the \HII~region is narrower
	and the resulting photoevaporation rate is larger by about a factor of two in \cite{2012_Owen}
	than those in our intermediate EX/Z0 with $f_h = 1$ (the right column in \fref{fig:X-raytest}).
	Since these differences likely originate from a number of 
	differences in the adopted methods,
	it would be difficult to specify the causes. 
	Nevertheless, 
	the results of our intermediate EX/Z0 with $f_h = 1$ and with the fiducial $f_h$ (\eqnref{eq:fh})
	provide important clues,
	because the different $f_h$ results in different opening angles
	and photoevaporation rates (\fref{fig:X-raytest} and \fref{fig:pratediscussion}).
	The results suggest that a higher heating efficiency $f_h$
	tends to generate a narrower \HII~region and a larger photoevaporation rate.
	Thus, we expect that 
	heating efficiency is higher in \cite{2012_Owen} than ours.	
	Actually,
	the temperature is typically $\sim 4000-5000\Kelvin$ 
	in the $2-10 \AU$ region in \cite{2012_Owen}, 
	while it is $2000-4000 \Kelvin$ in our intermediate EX/Z0 with $f_h = 1$. 
	Hence, we conclude that the narrower opening angle in \cite{2012_Owen} 
	may be attributed to a larger heating efficiency.
	The high heating efficiency could be 
	realized when the ionization degree is large
	owing to more efficient Coulomb interactions 
	between the ejected electron and the ambient electrons.
	We do not account for the effect in our simulations.
	This effect is neglected in the present study for simplicity.
	Incorporating the effect can yield
	a larger heating efficiency \citep{1985_ShullSteenberg},
	and X-ray heating rates could be increased 
	especially in the region where the ionization degree is high.
	On the other hand, since a highly ionized medium hardly absorbs X-rays, 
	the heating rate can be also lowered in such regions.
	With the electron abundances in the neutral regions of our model
	$\abn{e} \sim 10^{-4} - 10^{-2} $ (\fref{fig:theatcol} and \fref{fig:X-raytest}),
	taking account of the dependence on the electron abundance
	could increase X-ray heating rates by a small factor in the neutral regions.
	We note that a higher heating rate also 
	results in a narrower \HII~region in Run FEX and Run FE,
	where the opening angle is smaller when 
	FUV heating is strengthened by X-ray (\fref{fig:evaporations} and \fref{fig:eta}).

	\cite{2017_Wang} show that 
	disabling efficient cooling processes 
	results in large photoevaporation rates
	due to hard X-rays with $1\keV$.
	In order to examine if this is also the case in
	the fiducial model of the present study,
	we further perform
	a test simulation, 
	where $f_h = 1$ and the same thermal processes as Run X/Z0 are used 
	except that all of the line cooling are disabled.
	The resulting photoevaporation rate is $5\times 10^{-9} \myr$,
	which is modestly larger than our EUV photoevaporation rates.
	Thus, 
	if all of the primary electron energy go into heating
	and line cooling processes are not effective, 
	which might be unrealistic,
	hard X-rays can also cause a relatively efficient mass loss.

	Overall, 
	the effectiveness of X-rays on driving photoevaporation 
	significantly depends on the heating efficiency $f_h$
	and the exact shape of the spectrum of high energy photons especially around $\sim 0.1 \keV$. 
	Hence, for a comprehensive study of X-ray photoevaporation,
	it is necessary to model the heating efficiency  
	in a consistent manner,
	and to adopt a realistic spectrum.

        Finally, we note that the photoevaporation rate likely depends on  
	metallicity when assuming a hard EUV spectrum.
	Since the efficiency of radiative cooling due to metals in the neutral region, such as 
	\OI~cooling and dust-gas collisional cooling,
	decreases with metallicity, 
	there may be metallicity dependence 
	of EUV/X-ray photoevaporation
	when the exact spectral shape is taken into account.
	Similarly,
	different FUV spectra should also result in
	a different photoevaporation rate.
        Further studies are warranted to address these issues associated with detailed conditions.

	\subsection{Uncertainties in Input Parameters}
	We have concluded that 
	FUV effectively drives dense neutral flows,
	while X-ray is ineffective.
	The conclusion should depend on 
	input parameters such as 
	the abundance of polycyclic aromatic hydrocarbons (PAHs)
	and luminosities/spectra of FUV/X-ray.

	PAHs and very small grains significantly contribute 
	to photoelectric heating, and thus their abundances
	affect FUV photoevaporation rates
	\citep{2008_GortiHollenbach, 2009_Gorti, 2015_Gorti, 2018_Nakatani}. 
	Recent observations suggest that 
	the PAH abundances around T Tauri stars 
	might be smaller than the interstellar value,
	which we adopt in our fiducial model,
	although there are uncertainties 
	in the observational results \citep{2007_Geers, 2010_Oliveira, 2013_Vicente}. 
	In Paper I, we investigate 
	the effect of the reduced PAH abundance 
	on FUV photoevaporation. 
	We find that the net effect of the reduced PAH abundance is 
	weakening photoelectric heating,
	but FUV driven flows are excited anyway even without the PAH contribution.
	The resulting mass loss rates are the same orders of magnitude 
	as those with the PAH contribution to FUV heating.

	Besides the PAH abundance, 
	photoelectric heating rates depend on
	the local size distribution and amount of grains,
	which can be spatially variable in PPDs 
	owing to the effects of 
	dust growth, settling, and entrainment 
	\citep{2011_Owen_b, 2016_Hutchison_a, 2016_Hutchison}.
	Actually,  
	their variabilities are detected in several PPDs \citep[e.g.,][]{2016_Pinte}.
	Disk opacity is varied according to the spatial 
	distribution of dust grains,
	which also strongly affect
	photoevaporation rates. 
	Hence, in order to derive FUV photoevaporation rates accurately, 
	it is necessary to take account of different spatial distributions of grains with various sizes 
	as well as the reduced abundances of smaller grains.

	FUV and X-ray luminosities of young stars 
	are also uncertain factors. 
	A large fraction of FUV photons are considered to be produced within 
	accretion shocks around a classical T Tauri star (CTTS).
	CTTSs have a
	wide variety of FUV luminosities with $10^{-6}\, L_\odot \lesssim L_\text{FUV} \lesssim  L_\odot$
	which roughly correlates with the accretion rate $\dot{M}_\text{acc}$
	as $L_\text{FUV} \propto \dot{M}_\text{acc}$ \citep{1998_Gullbring,2012_Yang}.
	X-ray luminosities of T Tauri stars 
	range between $10^{28}\unit{erg}{}\unit{s}{-1} \lesssim L _\text{X} \lesssim 10^{31} \unit{erg}{}\unit{s}{-1}$
	\citep{2007_Gudel,2014_Vidotto},
	and measured plasma temperatures are typically $(5-30)\e{6} \Kelvin$ 
	which corresponds to 
	the peak X-ray energy of $\sim 1\keV$ \citep{2009_Gudel, 2014_Alexander}.
	A small number of T Tauri stars are known to have 
	``soft X-ray excess'' with temperatures of 
	a few million kelvin $(0.3-0.4\keV)$ \citep{2009_Gudel}. 
	TW Hya, whose X-ray spectrum is adopted in our fiducial model, is one of them.
	Thus, our fiducial X-ray spectra 
	might be relatively softer than a typical X-ray spectrum of a T Tauri star
	without the excessive soft X-ray.		
	Our results show that 
	X-ray is ineffective to drive photoevaporation 
	in our fiducial model 
	even though the fiducial spectrum contains the soft X-ray excess.
	Since softer components of a X-ray spectrum 
	play a major role in driving photoevaporative flows 
	(\secref{sec:speff} and \secref{sec:f_h}),
	using a typical X-ray spectrum without the soft X-ray excess
	would not change the conclusion
	that X-rays are ineffective to drive photoevaporation.
	We note that X-ray spectra constructed from emission measure data
	can be different from `processed' spectra that actually reach PPD surfaces, 
	considering the possibility of screening close to the X-ray source.
	We also note that the observed FUV/X-ray luminosities and spectra 
	have wide varieties as above; 
	FUV and X-ray luminosities independently vary with evolutional stages of PPDs.
	Hence, it is worth investigating relative importance of FUV/X-ray in photoevaporation 
	with a variety of luminosities and spectra.

\section{SUMMARY}	\label{sec:summary}
	
	We have performed a suite of radiation hydrodynamics simulations of 
	photoevaporating protoplanetary disks 
	to study the metallicity dependence of 
	photoevaporation due to FUV, EUV, and X-ray radiation.	
	Direct comparison between a variety of cases have shown that 
	X-rays alone do not heat disk gas up to sufficiently high temperature
	to cause a significant photoevaporative mass loss.
	
	Although the net heating effect is unimportant,
	X-rays effectively ionize the neutral region in a disk.
	Then the electron abundance in the neutral region
	is raised,
	and 
	charged dust grains recombine more efficiently.
	The FUV photoelectric heating efficiency is increased by the fast recombination,
	and the temperature in the neutral region becomes higher because of the strengthened 
	FUV heating.
	Consequently,
	including the
	X-ray radiation
	results in a larger photoevaporation rate,
	compared with the cases with FUV heating only.

	With FUV, EUV, and X-ray radiation, 
	the disk photoevaporation rate increases 
	as metallicity decreases in the range of $\metal \gtrsim 10^{-1.5} \, \smetal$
	because of the reduced opacity of a disk for FUV photons.
	At $\metal \lesssim 10^{-1.5}\, \smetal$,
	dust-gas collisional cooling becomes efficient compared to FUV photoelectric heating, 
	and suppress photoevaporation.
        In this metallicity range, 
	the strengthening effect of X-rays 
	is crucial to driving FUV photoevaporation.
	Without X-rays,
	the FUV heating does not excite photoevaporation
	and 
	only EUV-driven flows contribute to the mass loss.
	Therefore,
	the photoevaporation rate is
	significantly large in the simulations with very low metallicities 
	if the X-ray effects are incorporated.

	We derived the metallicity dependence 
	of the resulting photoevaporation rates.
	The metallicity dependence
	of photoevaporation rates due to FUV or strengthened FUV heating
	is consistent with 
	the observational metallicity dependence of 
	the disk lifetimes.
	Our model predicts that
	protoplanetary disks
	in an extremely low metallicity environment
	have longer lifetimes 
	than in solar or sub-solar metallicity environments.

\acknowledgments
We thank Neal Turner, Mario Flock, Uma Gorti,  Shu-ichiro Inutsuka, Kengo Tomida, and Kei Tanaka
for fruitful discussions and helpful comments on the paper.
We also thank the
anonymous referee for giving insightful comments to improve
the manuscript.
RN has been supported by the Grant-in-aid for the Japan Society 
for the Promotion of Science (16J03534) and by Advanced Leading 
Graduate Course for Photon Science (ALPS) of the University of Tokyo.
TH appreciates the financial supports by the Grants-in-Aid for Basic Research by the Ministry of Education, Science and Culture of Japan (16H05996).
HN appreciates the financial supports by Grants-in-Aid for Scientific Research (25400229).
RK acknowledges financial support via the Emmy Noether Research Group on Accretion Flows and Feedback in Realistic Models of Massive Star Formation 
funded by the German Research Foundation (DFG) under grant no. KU 2849/3-1.
All the numerical computations were carried out on Cray XC30 and Cray XC50
at Center for Computational Astrophysics, National Astronomical Observatory of Japan.

\bibliography{template}
\bibliographystyle{apj}

\appendix
\section{Implementation of X-ray heating/ionization}
\label{app:X-ray}
	\subsection{Cross-Section}
	In general, X-rays are absorbed by both non-metal and metal elements
	which compose both gas and dust grains in a medium.
	The cross section of the dust grains for X-rays typically 
	contributes relatively small to the total cross section of 
	the medium \citep{2000_Wilms}.
	Hence, we ignore the contribution of the dust grains
	to the cross section for X-rays in this study.

	We use the total cross section presented by \cite{2004_Gorti} (hereafter, GH04).
	GH04 calculates the total cross section
	for a solar metallicity disk
	and gets a fitted total cross section
	per hydrogen nuclei
	\begin{equation}
		\crsc{GH04}(E) = 1.2\e{-22} \braket{\frac{E}{1\keV}}^{-2.594}~\cm{2},	\label{eq:GH04_eqB3}
	\end{equation}
	where $E$ is the X-ray photon energy.
	As metallicity increases (decreases),
	the contribution of the metal elements to
	\eqnref{eq:GH04_eqB3} 
	increases (decreases) 
	in the energy range of $E \geq 0.29\keV$,
	where $0.29\keV$ corresponds to 
	the threshold energy for carbon ionization \citep{2000_Wilms}.
	Therefore, 
	we modify \eqnref{eq:GH04_eqB3}
	to a metallicity-dependent cross section
	\begin{eqnarray}
		&\crsc{H}  &= 11.55\e{-24} \braket{\frac{E}{1\keV}}^{-3.4} 	 \cm{2}, \label{eq:crssec_HHe}	\\
		&\crsc{}	&=
		\left\{
		\begin{array}{l l}
			\crsc{H}
			& ~~~\left(0.1\keV\leq E \leq 0.29\keV \right)     \\
			{\rm max} \left(
			\crsc{H},
			~\crsc{GH04} \times \dfrac{\metal}{\smetal} \right)~\cm{2}  		
			&   ~~~\left( E \geq 0.29\keV \right)
		\end{array}
		\right.
		,	
		\label{eq:crosssection_ours}
	\end{eqnarray}
	where $\crsc{H}$ is the hydrogen cross section per hydrogen nuclei
	presented by GH04.

	The ionization cross section of helium is about four times larger 
	than that of hydrogen \citep{1985_ShullSteenberg,2004_Gorti},
	and thus helium contributes to X-ray absorption.
	However, 
	since helium abundance 
	is typically ten times smaller than hydrogen,
	including helium contribution to \eqnref{eq:crosssection_ours}
	raises the absorption rate 
	by only $\lesssim 40\%$.
	Therefore, in our chemistry model,
	we neglect the helium contribution to X-ray absorption
	in order to save computational cost for calculating the chemistry
	and use \eqnref{eq:crosssection_ours}
	as the cross section for X-rays.

	\subsection{Spectral Energy Distribution of X-rays}
	For the spectral energy distribution (SED)
	of X-rays, we use the SED presented in \cite{2007_Nomura_II}.
	The SED is that of TW Hya which is one of the classical T Tauri stars.
	We set the minimum energy of the X-rays
	to $\Emin = 0.1\keV$
	and the maximum energy of it
	to $\Emax = 10\keV$.
	The absolute values 
	of the SED are normalized 
	so that the total X-ray luminosity
	is calculated to be $L_\text{X} = 1\e{30}~\unit{erg}{}~\unit{s}{-1}$.

	\subsection{X-ray Ionization}
	When X-rays are absorbed by elements,
	photoelectrons are ejected from the absorbers.
	The photoelectrons are called primary electrons.
	The energy of the primary electrons 
	is so high that they further ionize the ambient neutral gas.
	This is so-called secondary ionization 
	and the ejected electrons by secondary ionization
	are called secondary electrons.
	
	In this study,
	we implement the X-ray ionization 
	of atomic and molecular hydrogen,
	who are the most abundant species.
	The total number of the primary electrons per unit volume is 
	\begin{equation}
		N_\text{prim}
		= \int_{E_\text{min}}^{E_\text{max}} 	
		dE ~ \crsc{} \nh \frac{F(E)}{E} e^{-\taux} ,
	\end{equation}
	where $\taux$ is the optical depth for X-rays
	defined as $\taux = \crsc{}\col{H}$.
	For atomic hydrogen, 
	the amount of energy which goes to secondary ionization
	is $\Phih (E-E_\text{th})$, where
	$\Phih$ is the fraction of primary electron energy
	consumed by secondary ionization and
	$E_\text{th}$ is the threshold energy of the ejector of a primary electron.
	The threshold energy is assumed to typically be 
	much smaller than the ejected 
	non-thermal electron energy.
	In that case,
	the number of secondary ionizations can be
	calculated as $\Phih E / 13.6 \eV $.	
	Consequently, 
	the {\it total} number of secondary 
	ionizations per unit volume is
	\begin{equation}
		\secondioni =  \int_{\Emin}^{\Emax} 	
		dE ~ \frac{F(E)}{E} e^{-\taux}~  \crsc{} \nh	~
		\braket{\frac{\Phih E}{13.6\eV}},
		\label{eq:second}
	\end{equation}	
	while the total number of photoionization per unit volume is 
	\begin{equation}
		\primaryioni =  \int_{\Emin}^{\Emax} 	
		dE ~ \frac{F(E)}{E} e^{-\taux}~  \crsc{H} ~  \nspe {HI} ~.
	\end{equation}		
	Hence, 
	the total ionization rate of atomic hydrogen by X-rays is 
	\begin{eqnarray}
		R_\text{X,H} 	&&=	\braket{\primaryioni + \secondioni}/\nh \nonumber \\
					&&= \int_{E_{\rm min}}^{E_{\rm max}} dE 
					~ \frac{F(E)}{E} e^{-\taux} 
					\left[\crsc{H}~ \abn{HI} 
					+ \crsc{} \braket{\frac{\Phih E}{13.6\eV}}
					\right] .
					\label{eq:totioni}
	\end{eqnarray}
	
	In general,
	the energy fraction consumed by secondary ionization $\Phih$ 
	depends on electron abundance and primary electron energy
	\citep{1996_Maloney},
	but it is usually simplified as the function 
	which depends only on electron abundance.
	When electron abundance is sufficiently low such as 
	$\abn{e} \lesssim 0.01$,
	approximately 35\% of the primary electron 
	energy goes into secondary ionization 
	and thus $\Phih$ can be approximated by $\Phih \simeq 0.35$
	 \citep{1996_Maloney, 2004_Gorti}.
	In a protoplanetary disk, even if X-rays ionize the medium,
	electron abundance is typically low $\abn{e} \ll 1$
	except the \HII~region where EUV photons ionize the gas.
	In the \HII~region, 
	the electron abundance is $\abn{e} \sim 1$ 
	and the atomic hydrogens are almost completely ionized $(\abn{HI}\sim0)$.
	The approximation $\Phih \simeq 0.35$
	is not appropriate in this case.
	As electron abundance increases,
	a smaller amount of energy goes into secondary ionization.
	In order to simply incorporate this effect,
	we use 
	\begin{equation}
		\Phih	=	0.35 \abn{HI},	\label{eq:probabilityphih}
	\end{equation}
	as a first approximation.
	Hence, \eqnref{eq:totioni} reduces to  
	\begin{equation}
		R_\text{X,H} = \abn{HI} \int_{\Emin}^{\Emax} dE ~ \frac{F(E)}{E} 
		e^{-\taux} \left[\crsc{H} + \crsc{} \braket{\frac{25.7E}{1\keV}}\right] .
		\label{eq:rxh}
	\end{equation}

	For the X-ray ionization of molecular hydrogen,
	we use $2\crsc{H}$ as the cross section of molecular hydrogen
	as GH04
	\citep[cf.][]{1998_Yan,  2011_Draine}.
	Note that the ratio of the cross section of a hydrogen molecule 
	to that of a hydrogen atom 
	is $\sim 2.8$ in the high energy limit 
	\citep{2000_Wilms,2001_YanErratum, 2011_Draine}. 
	The number of primary ionization for molecular hydrogens is 
	\begin{equation}
		\primary{\ce{H2}} =  \int_{\Emin}^{\Emax} 
			dE ~ \frac{F(E)}{E} e^{-\taux}~  (2\crsc{H}) ~  \nspe {\ce{H2}} ~.
	\end{equation}
	Regarding the secondary ionization,
	the fraction of primary electron energy going into secondary ionization 
	is similar to that of atomic hydrogens 	
	if electron abundance is small
	\citep{1996_Maloney}.
	Thus, $\Phih$ is used also for secondary ionization of molecular hydrogens.
	The number of the secondary ionization of \ce{H2} per photoelectron is
	changed from \eqnref{eq:second}
	by the difference between the ionization energies 
	of an atomic and molecular hydrogen.
	We calculate the secondary ionization of molecular hydrogens as
	\begin{eqnarray}
		\second{\ce{H2}} = 2\abn{\ce{H2}}  \int_{\Emin}^{\Emax} &&dE ~ 
		\frac{F(E)}{E} e^{-\taux}~  \crsc{} \nh	~ \nonumber \\
				&&\times \braket{\frac{13.6\eV}{15.4\eV}}\braket{\frac{25.7E}{1\keV}} .
				\label{eq:h2ionize}
	\end{eqnarray}
	The first factor $2\abn{\ce{H2}}$ is used instead of $\abn{HI}$ in \eqnref{eq:probabilityphih}.
	Since the maximum value of $\abn{\ce{H2}}$ is 0.5,
	the factor becomes unity in \eqnref{eq:h2ionize}
	in fully molecular gas.
	After all, the total ionization rate of molecular hydrogens reduces to
	 \begin{eqnarray}
	 R_\text{X,\ce{H2}} = 
	  &&\braket{ \primary{\ce{H2}} + \second{\ce{H2}}}/\nh \nonumber 	\\
	  =&&	2\abn{\ce{H2}} \int_{\Emin}^{\Emax} dE ~ \frac{F(E)}{E} 
		e^{-\taux} \left[\crsc{H} + \crsc{} \braket{\frac{22.7E}{1\keV}}\right] .
		\label{eq:rxh2}
	\end{eqnarray}

	X-rays also ionize helium by primary and secondary ionization.
	The fraction of primary electron energy consumed by secondary ionization of helium
	is $\Phi_\text{He}\sim 0.05$ \citep{1985_ShullSteenberg,2004_Gorti},
	which is much less than that of hydrogen.
	Taking into account this weaker secondary ionization
	and the fact that helium abundance 
	is an order of magnitude less than hydrogen abundance, 
	helium ionization rates would be smaller than those of hydrogen.
	Since helium recombination rates 
	and hydrogen recombination rates
	have only a small difference
	\citep[about a factor of two in the range of $10\Kelvin \leq T \leq 1000\Kelvin$, ]
	[]{2013_McElroy},
	the ionization degree of helium due to X-rays would be smaller than or 
	at least the same order of magnitude as 
	hydrogen ionization degree due to X-rays.
	Thus, including X-ray ionization of helium in our chemistry
	might raise electron abundance in the neutral region of a disk
	by several tens of percent.
	In the sense that FUV heating is strengthened more 
	by the higher electron abundance as discussed in \secref{sec:result1}, \secref{sec:result2},
	and \secref{sec:result3},
	taking account of helium ionization due to X-ray would make our conclusion firmer.
	Nevertheless, 
	it is expected that including helium ionization 
	does not significantly affect our results,
	and so
	we do not incorporate helium ionization in our chemistry
	to save the computational cost for the calculation of the chemistry network.

	\subsection{X-ray Heating}	\label{sec:xheat}

	The total energy deposited to primary electrons by X-ray absorption
	is given by
	\begin{equation}
		\epsilon_\text{tot} = \int_{\Emin}^{\Emax} dE ~ F(E) e^{-\taux}  \crsc{} \nh 	.
	\end{equation}
	Not all the energy 
	goes into heating 
	due to other processes such as 
	secondary ionization and excitations of the ambient gas.
	If hydrogens are purely atomic in gas,
	about $10\%$ of the deposited energy by the absorption of X-rays
	results in gas heating \citep{1996_Maloney}.
	If hydrogens are purely molecular,
	about $40\%$ of the deposited energy
	thermalizes.
	Therefore, in this study,
	we set the X-ray heating rate to
	\begin{eqnarray}
		\Gamma_\text{X} &&= ~f_h ~\epsilon_\text{tot},	 \label{eq:X-rayheat}\\
		f_h  && \equiv
		\frac {0.1 \abn{HI} + 0.4 \abn{\ce{H2}} }{\abn{HI} + \abn{\ce{H2}}},	\label{eq:fh}
	\end{eqnarray}
	where $f_h$ is the heating efficiency of the X-ray heating rate.

\section{Chemical Reactions}
	We use the same chemistry network as that of Paper I
	other than 
	we add chemical reactions 
	involving \ce{H2+} to the network.
	The added chemical reactions are
	listed in \tref{tab:chem_reac}.

\begin{table*}
\begin{center}
Table B1. Chemical reactions incorporated additionally in our simulations \\[3mm]
\begin{tabular}{lllc}\hline
Label   & Reaction      & Rate Coefficient & Reference \\ \hline \hline
k9	& \ce{H + H+ -> H2+  } 	& $10^{ -19.38 - 1.523  \log T + 1.118 (\log T) ^2 - 0.1269  (\log T)^3}$ & 1\\ 
k10 	& \ce{H2+ + H -> H2 + H+}	& $6.4\e{-10}$      	& 1\\
k11     & \ce{H2 + H+  -> H + H2+}  	& $3.0\e{-10}\exp[-2.1050\e{4}\Kelvin/T]$                                     	& 1\\
p4	&\ce{H} + $\gamma$ \ce{-> H+ + e}	& $R_\text{X,H}$ ~(cf. \eqnref{eq:rxh})& 2     \\
p5    	&\ce{H2} + $\gamma$ \ce{-> H2+ + e} & $R_\text{X,\ce{H2}}$ ~(cf. \eqnref{eq:rxh2})& 2     \\
\hline
\end{tabular}
\end{center}
Reference ----- (1) \cite{2000_Omukai}~~~       (2) \cite{2004_Gorti}
\label{tab:chem_reac}
\end{table*}

\end{document}